\documentclass[lettersize,journal]{IEEEtran}
\usepackage{cite}
\usepackage{amsmath,amssymb,amsfonts,amstext}
\usepackage{graphicx}
\usepackage{textcomp}
\usepackage[utf8]{inputenc}
\usepackage{balance}
\usepackage{comment}
\usepackage{soul}
\usepackage{subcaption}

\usepackage{mathtools}
\usepackage{float}

\usepackage{amsthm}

\newtheorem{definition}{Definition}

\newtheorem{remark}{Remark}

\newtheorem{problem}{Problem}

\usepackage{algorithm}
\usepackage{algorithmicx} 
\usepackage{algpseudocode} %

\DeclareMathOperator*{\minimize}{Minimize}
\DeclareMathOperator*{\maximize}{Maximize}

\usepackage{todonotes}
\def\BibTeX{{\rm B\kern-.05em{\sc i\kern-.025em b}\kern-.08em
    T\kern-.1667em\lower.7ex\hbox{E}\kern-.125emX}}
\usepackage{balance}

\makeatletter
\let\NAT@parse\undefined
\makeatother
\usepackage{hyperref}
\hypersetup{
  colorlinks=true,
    linkcolor= blue,
    allcolors=blue,
    citecolor = blue,
    filecolor=black,      
    urlcolor=blue,
    }

\begin{document}
\title{On Mobility Equity and the Promise of Emerging Transportation Systems}

\author{Heeseung Bang, 
Aditya Dave, 
Filippos N. Tzortzoglou, 
Shanting Wang, \\and Andreas A. Malikopoulos, \textit{Senior Member, IEEE}
    \thanks{This research was supported partially by NSF under Grants CNS-2149520, CMMI-2219761, by Mathworks, and by Delaware Department of Transportation.}
    \thanks{The authors are with the School of Civil and Environmental Engineering, Cornell University, Ithaca, NY 14850, USA. {\tt\small email: \{h.bang,a.dave,ft253,sw997,amaliko\}@cornell.edu}}
}


\maketitle

\begin{abstract}
This paper introduces a mobility equity metric (MEM) for evaluating fairness and accessibility in multi-modal intelligent transportation systems.
The MEM simultaneously accounts for service accessibility and transportation costs across different modes of transportation and social demographics.
We provide a data-driven validation of the proposed MEM to characterize the impact of various parameters in the metric across cities in the U.S.
We subsequently develop a routing framework that aims to optimize MEM within a transportation network containing both public transit and private vehicles.
Within this framework, a system planner provides routing suggestions to vehicles across all modes of transportation to maximize MEM.
We evaluate our approach through numerical simulations, analyzing the impact of travel demands and compliance of private vehicles. This work provides insights into designing transportation systems that are not only efficient but also equitable, ensuring fair access to essential services across diverse populations.
\end{abstract}



\section{Introduction}

Rapid urbanization and population growth over the last few decades have led to increasing transportation demands across cities worldwide. Although transportation infrastructure has expanded to meet these needs, the expansion is ultimately constrained by limited space and the challenges of reconstructing existing infrastructure. 
Consequently, transportation efficiency has not kept pace with the increasing demands, creating a critical challenge in ensuring continued access to essential goods and services for urban populations.

Emerging mobility systems, such as connected and automated vehicles (CAVs), shared mobility, and on-demand mobility systems, have gained significant attention as promising solutions to reduce traffic congestion, operation costs, and environmental impact in urban transportation networks.
For instance, at the vehicular control level, recent research has focused on optimizing energy efficiency in CAVs in different traffic scenarios such as lane-changing \cite{wang2019q}, merging on-ramps in mixed traffic \cite{Nishanth2023AISmerging}, signalized \cite{genser2024time,mahbub2022_ifac} and unsignalized single intersections \cite{Malikopoulos2020} and adjacent intersections \cite{chalaki2020TITS},  roundabouts \cite{chalaki2020experimental}, and corridors \cite{Zhao2018ITSC}.
These efforts have also extended to the network level and addressed vehicle routing problems. For example, some approaches have combined efficient routing with coordination strategies \cite{Bang2023flowbased,Bang2022combined} or utilize learned traveler preferences to achieve social objectives \cite{biyik2019green}.

Despite considerable efforts in developing and optimizing emerging mobility systems, effective strategies that address societal challenges, such as mobility equity, remain needed. While some studies have examined equity issues within public transit, the exploration of equity within the context of intelligent transportation systems remains an open question. In this paper, we provide an analysis of this question from the perspective of combining equity with optimal routing.

\subsection{Related Work}

\subsubsection{Accessibility}
The study of mobility equity has its roots in measuring accessibility to various modes of transportation and opportunities within a transportation network \cite{hansen1959accessibility,cervero1997paradigm}. Accessibility for an individual within a transportation network is a multi-faceted concept capturing the user's ability to access various opportunities and essential services, thus achieving their travel demands \cite{martens2016transport, Kostopoulou2024}.

A majority of research efforts have focused on measuring accessibility to \textit{jobs} for travelers hailing from different demographics within a city. For instance, Curl \cite{curl2018importance} proposed a simple measure of accessibility based purely on geographical distance to a job from a traveler's starting location. 
More sophisticated measures also take into account the travel time within a transportation network, such as \textit{isochronic measures} \cite{cervero2005accessible,o2000using,fan2012impact,golub2014using,tian2017effects}, which estimate reachable geographical regions within a certain time threshold using the transportation network and utilize a count of the number of jobs or other opportunities (such as education) as a representation of accessibility.
To measure the accessibility afforded to underprivileged travelers, Deboosere and El-Geneidy \cite{deboosere2018evaluating} propose the inclusion of only low-income jobs in their isochronic measure.
Mavoa et al. \cite{mavoa2012gis} introduced an accessibility index that explicitly accounts for the time taken to walk to a public transit station and waiting times at stations. 
Cui et al. \cite{cui2019accessibility} utilized isochronic measures to examine the relationship between accessibility and commute duration for low and high-income individuals.
A contrasting approach to account for travel time utilizes \textit{gravity-based measures}. These measures interpret travel time as a disutility incurred by a traveler and, thus, consider the average travel time to access all opportunities within the network as a measure of accessibility \cite{el2007mapping,manaugh2012benefits,grengs2010intermetropolitan}. A modification to gravity-based measures was proposed by
Guzman et al. \cite{guzman2017assessing} to consider both travel time and associated fare of travel for a joint disutility.
El-Geneidy et al. \cite{el2016cost} showed that combining travel time with transit fares yields a better measure of accessibility by accounting for the disadvantages of low-income populations. 

\subsubsection{Mobility Equity}

While the aforementioned studies on accessibility provide valuable insights into the overall level of access to opportunities, they often fail to fully capture the disparities across different regions and groups.
To address this limitation, researchers have increasingly focused on the concept of equity, which refers to the fair distribution of resources and benefits across society. In particular, mobility equity goes beyond quantifying accessibility and aims to provide uniform access to transportation services and opportunities to all individuals within a network, regardless of their location or socio-economic status.

In the studies on equity in transportation, two primary dimensions are widely recognized: vertical and horizontal equity. Horizontal equity focuses on providing equitable treatment to individuals with similar transportation needs, irrespective of their socio-economic status or geographic location \cite{mortazavi2017framework,chan2021evaluating}. For example, Brocker et al. \cite{brocker2010assessing} developed a spatial equilibrium model to estimate the spatial equity in welfare effects of investment into transportation infrastructure.
Jang et al. \cite{jang2017assessing} characterized the impact of planned network expansions, e.g., the addition of subway lines, on accessibility equity across the city of Seoul.
However, only accounting for horizontal equity fails to consider differences between different societal groups.

Vertical equity, on the other hand, ensures fair access to transportation services for individuals from different socio-economic backgrounds. This includes measuring income-based disparities in transportation access, mitigating the financial burdens of transportation costs on low-income households, and enhancing accessibility for marginalized communities, such as individuals with disabilities or the elderly \cite{calafiore202220,el2016cost,ricciardi2015exploring,foth2013towards,lee2023assessing}.
In most of these studies, a variant of the Gini index is used to measure inequality in the distribution of resources  \cite{gastwirth1972estimation,pritchard2019potential}.
For example, Mortazavi and Akbarzadeh \cite{mortazavi2017framework} proposed accessibility to public transit for car-less populations and used the Gini index to characterize equity in access across different demographics.
Similarly, Delbosc and Currie \cite{delbosc2011using} used the Gini index to evaluate the equity in access to public transit across districts in Melbourne.
Jin et al. \cite{jin2019uber} also used the Gini index to analyze urban transportation equity for each mode, such as ride-hailing, taxi, and public transit.

These equity measures allow us to identify areas where improved service or transit infrastructure would maximize accessibility benefits for disadvantaged populations \cite{karner2018assessing}.
However, improving service or transit infrastructure is a task for urban planning, requiring large stretches of unused space within the city and incurring high monetary/social costs.
Thus, it is important to study how mobility equity may be improved without changing the infrastructure of a city.
One promising approach to achieve this goal is to utilize emerging mobility systems to improve traffic congestion. This research direction suffers from a lack of consensus on how mobility equity should be defined in the context of vehicle routing \cite{bills2017looking,brown2022aspiration}, since there is no universally adopted method to achieve mobility equity in such systems \cite{litman2017evaluating,ferrell2023defining}. In fact, existing methods do not account for the complex dynamics within emerging mobility systems. 
For example, Cohn et al. \cite{cohn2019examining} investigate the potential impacts of introducing autonomous vehicles (AVs) on transportation equity in the Washington, D.C. area. Still, their results hinge upon their AV modeling assumptions, which oversimplify the complexities of AV deployment and usage.
Because of this gap, a more holistic approach is required to achieve equity in the operation of emerging mobility systems.


\subsubsection{Emerging Mobility Systems}

The advent of disruptive technologies such as autonomous driving, connectivity, and power-train electrification has enabled the development of new mobility paradigms, such as autonomous mobility-on-demand (AMoD) systems, where self-driving vehicles provide on-demand transportation services \cite{zhao2019enhanced}. Several approaches have been proposed to study and optimize the operation of AMoD systems, including agent-based simulation \cite{levin2017general}, queueing-theoretical models \cite{iglesias2019bcmp}, and network flow-based models \cite{rossi2018routing}.

In particular, multi-commodity network flow models have emerged as a powerful tool for AMoD system optimization, as they allow for consideration of multiple objectives, avoid scaling with fleet size, and can capture operational constraints. These models have been successfully applied to study a range of problems, such as congestion-aware routing \cite{salazar2019congestion}, ride-pooling \cite{tsao2019model}, electric vehicle fleets \cite{luke2021joint,bang2021AEMoD}, vehicle-to-grid interactions \cite{rossi2019interaction,turan2019smart,estandia2021interaction}, and intermodal transportation \cite{salazar2020intermodal,wollenstein2021routing}.
However, despite the significant research efforts in optimizing AMoD system operations, a vast majority of the studies have focused on minimizing 
metrics such as travel time or operational costs \cite{spieser2014toward}. While these metrics capture important economic indicators, they fail to account for transportation equity.
As AMoD systems and other emerging mobility solutions become increasingly prevalent, it is crucial to infuse optimization frameworks with social and sustainability perspectives to ensure these technologies truly benefit society as a whole \cite{sheller2018theorising,servou2023data}.

Our previous work \cite{Bang2023mem,bang2024cts} laid the foundation for the current study, exploring various aspects of mobility equity in intelligent transportation systems. 
We introduced a preliminary mobility equity metric using average accessibility across neighborhoods \cite{Bang2023mem}, and proposed an improved metric addressing earlier limitations \cite{bang2024cts}.
Similarly, Salazar et al. \cite{salazar2024accessibility} evaluated weighted-average unfairness and solved a routing problem, providing additional perspectives on quantifying and addressing mobility inequities in transportation networks and further emphasizing the importance of this research direction.

\subsection{Contributions}

In this paper, we propose an enhanced mobility equity metric (MEM) and an equity-focused routing framework for emerging mobility systems.
Our approach begins by introducing a \textit{mobility index} (MI) evaluating the accessibility to essential services from a specific node in the transportation network. The goal of the MI is to incorporate various factors such as travel time, travel costs, the importance of different services, and other demographic features.
We capture the impact of travel costs in the index using a gravity-based approach. Meanwhile, we quantify accessibility based on the number of reachable services within specified time constraints from selected origins. Importantly, this method relies solely on publicly available data, and we present a method of measuring MI with isochrone and point-of-interest (POI) data \cite{xi2018exploring}.
To assess the equitable distribution of MIs across a city, we employ the Gini index, a widely used measure of inequality. We conduct a comprehensive evaluation of MIs across $12$ major cities in the United States and analyze the current equity in those cities based on our MI definition. Additionally, we provide an analysis of the MEM to the choice of different parameters using real data from U.S. cities.

Furthermore, we introduce a ``system planner" designed to optimize route suggestions for vehicles within the network to enhance MEM. Notably, our system accommodates private vehicles, which may or may not adhere to the planner's recommendations, reflecting real-world transportation dynamics. The existence of non-compliant vehicles naturally forms the routing game between system planners and non-compliant vehicles.
In studies involving routing games, the Wardrop equilibrium model has been commonly utilized to predict traffic patterns. However, this model assumes that all private vehicles select their optimal routes, which may not always hold in practice. Hence, we adopt the cognitive hierarchy model, which considers different decision-making capabilities among drivers. We compare traffic patterns for different scenarios with different maximum levels of cognition and demonstrate that they eventually converge to the Wardrop equilibrium as the maximum level approaches infinity.

The primary contributions of this paper can be summarized as follows.
\textit{1) Introduction of MEM:} We introduce a novel MEM that evaluates the distribution of the \textit{ability to move} among different regions considering societal factors. Additionally, we evaluate our MEM in different cities to analyze and provide insights into the mobility equity of actual U.S. cities.
\textit{2) Development of a system planner:} We present a method to improve MEM within a multi-modal emerging mobility system using a system planner. This planner offers route suggestions aimed at enhancing mobility equity across various regions.
\textit{3) Analysis of the system planner's impact on MEM:} Through numerical simulations, we analyze the impact of the system planner's route suggestions on MEM. We utilize POI data and a traffic network in Boston, MA, USA. The resulting analysis provides insights into the effectiveness and implications of our MEM and routing approach in practical scenarios.

The remainder of this paper is organized as follows.
In Section \ref{sec:mem}, we introduce the MEM and explain a method to evaluate it using real data. In Section \ref{sec:routing}, we formulate a routing problem and incorporate it with MEM optimization problem.
We conduct numerical simulations and analyze results in Section \ref{sec:simulation}, and finally, in Section \ref{sec:conclusion}, we discuss concluding remarks and possible future work.


\section{Evaluation of Mobility Equity} \label{sec:mem}

In this section, we present our mobility equity metric (MEM) for transportation networks. To evaluate this equity, mobility from different regions in the network must be analyzed while considering socio-economic factors such as travelers' financial demographics, origins and destinations, and reasons for travel.
Additionally, features of the provided mobility services should be considered, i.e., available transportation modes, associated costs, and overall travel time. Our MEM simultaneously addresses these diverse factors across a transportation network. After presenting this metric, we also evaluate the mobility equity across several major U.S. cities using real-world data with our MEM.

\subsection{Transportation Network and Mobility}
We model the transportation network as a directed graph $\mathcal{G} = (\mathcal{V},\mathcal{E})$, where $\mathcal{V}\subset\mathbb{N}$ is the set of nodes and $\mathcal{E}\subset\mathcal{V}\times\mathcal{V}$ is the set of edges. The set $\mathcal{M}$ represents various modes of transportation available to travelers, e.g., public transportation, shared mobility, private vehicles, cycling, and walking. The set $\mathcal{S}$ contains the different types of services accessible through the transportation network.
To measure mobility equity, we focus on travel related to accessing essential services within the network, such as medical facilities, financial services, groceries, schools, entertainment venues, and restaurants. The choice of service types can be adapted to specialize or generalize our notion of equity in different applications.

For each mode $m\in\mathcal{M}$,  we define $c_m$ as the cost per passenger mile.
For each service type $s \in \mathcal{S}$, we define $\beta^s$ as the priority level of services type $s$, where a higher value indicates greater importance. 
This allows us to prioritize accessibility to essential services, such as medical facilities, groceries, financial services, and schools, over non-essential services, such as entertainment venues and shopping malls, when quantifying mobility equity.
Furthermore, the time threshold $\tau_m$ represents a reasonable limit on the travel time to access services using mode $m$.
For a node $i \in \mathcal{V}$, $\sigma_{i,m}^s(\tau_m)$ represents the count of services of type $s \in \mathcal{S}$ accessible within time $\tau_m$ using mode $m$ from node $i$.
These terms constitute a mobility index (MI) that measures accessibility from any node $i \in \mathcal{V}$.

\begin{definition} \label{def:mobility_index}
\vspace{8pt}
For a given network $\mathcal{G}$ with a set of modes $\mathcal{M}$ and services $\mathcal{S}$, the mobility index (MI) at node $i\in\mathcal{V}$ is
\begin{equation} \label{eq:MEM}
    \varepsilon_i = \sum_{m\in\mathcal{M}} e^{-\kappa_i c_m} \cdot \left\{ \sum_{s\in\mathcal{S}} \beta^s \tilde{\sigma}_{i,m}^s(\tau_m) \right\},
\end{equation}
where $\kappa_i$ is a constant of price sensitivity, capturing the importance allocated to travel cost over travel time, and $\tilde{\sigma}_{i,m}^s(\tau_m)$ is the normalized number of accessible services within the time threshold $\tau_m$ using mode $m \in \mathcal{M}$.
\end{definition}

In Definition \ref{def:mobility_index}, we normalize the number of accessible services from any node $i$ using mode $m$ with the \textit{maximum} number of accessible services across all nodes in $\mathcal{V}$ with mode $m$, i.e.,
\begin{equation}
    \tilde{\sigma}_{i,m}^s(\tau_m) = \frac{\sigma_{i,m}^s(\tau_m)}{\max_{v\in\mathcal{V}} \sigma_{v,m}^s(\tau_m)}.
\end{equation}
This normalization reduces the impact of an unbalanced number of accessible services across different service types. If we used the count $\sigma_{i,m}^s(\tau_m)$ to compute the MI in \eqref{eq:MEM} instead of $\tilde{\sigma}_{i,m}^s(\tau_m)$, we would be susceptible to the number of restaurants and cafes dominating the MI over sparsely distributed services such as hospitals, even if the priority levels were set differently.

The MI introduced in Definition \ref{def:mobility_index} incorporates travel time, user cost, price sensitivity, transportation modes, and service types into one measure of accessibility for node $i \in \mathcal{V}$.
By weighing services based on priority, we ensure that our metric is robust to the inclusion and exclusion of various services or, more generally, to various travel purposes.
Furthermore, we allow the price sensitivity $\kappa_i$ to vary with the choice of node $i \in \mathcal{V}$ to account for demographic differences in the ability to pay.
This is because a node $i$ within a transportation network typically represents a certain neighborhood in the underlying geography. Thus, each node $i\in\mathcal{V}$ may evaluate the trade-off between travel time and travel cost differently.
We can capture the impact of different demographics in the MI by appropriately selecting $\kappa_i$ for each $i \in \mathcal{V}$.
For example, individuals originating from a poor neighborhood $i$ may have a lower ability to pay for private vehicles, and thus, this neighborhood would have a higher value of $\kappa_i$.
Note that in neighborhoods with high price sensitivity, the MI rewards an increase in accessibility of services using low-cost modes of transportation, such as public transit, over high-cost modes, such as private vehicles. Thus, we can understand the MI $\varepsilon_i$ as a good measure for the \textit{ability to move} of travelers originating from any node $i \in \mathcal{V}$.

\begin{figure}
    \centering
    \includegraphics[width=0.8\linewidth]{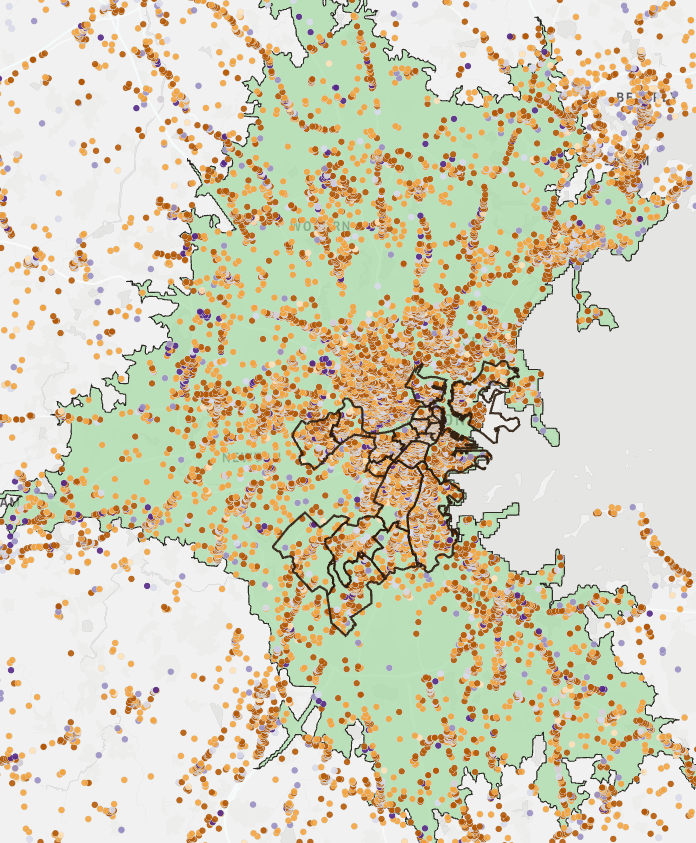}
    \caption{Neighborhoods in Boston, MA, USA, and service locations. The green area depicts the accessible distance within 30 minutes of free-flow driving from the city center.}
    \label{fig:neighborhoods}
\end{figure}

\subsection{Mobility Index Evaluation}

To evaluate the MI in Definition \ref{def:mobility_index} for any node $i \in \mathcal{V}$ within a transportation network, we must first compute the number of accessible services $\sigma_{i,m}^{s}(\tau_m)$ of each type $s \in \mathcal{S}$ using each mode $m \in \mathcal{M}$. 
To achieve this, we leverage publicly available isochrone data.
Each isochrone is a polygon surrounding a node $i$ that captures the boundary of the geographic area that can be accessed using mode $m$ within a time threshold $\tau_m$.

\begin{figure}
    \centering
    \subfloat[]{\includegraphics[width=0.45\linewidth]{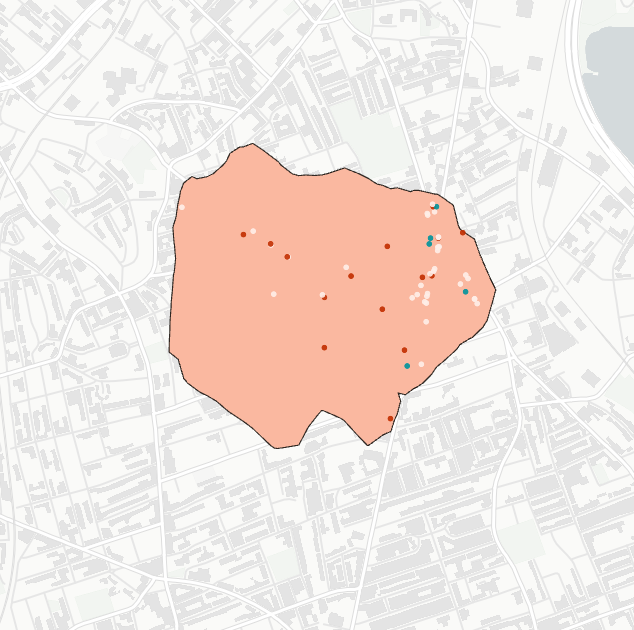}}
    \subfloat[]{\includegraphics[width=0.45\linewidth]{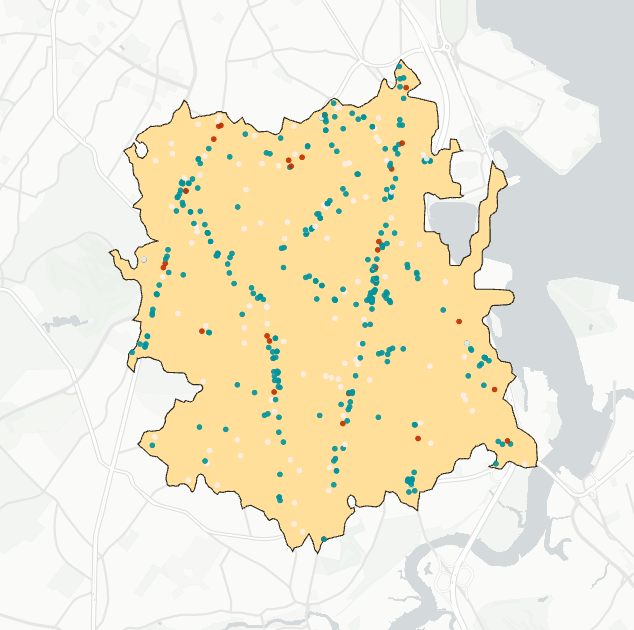}}\\
    \subfloat[]{\includegraphics[width=0.45\linewidth]{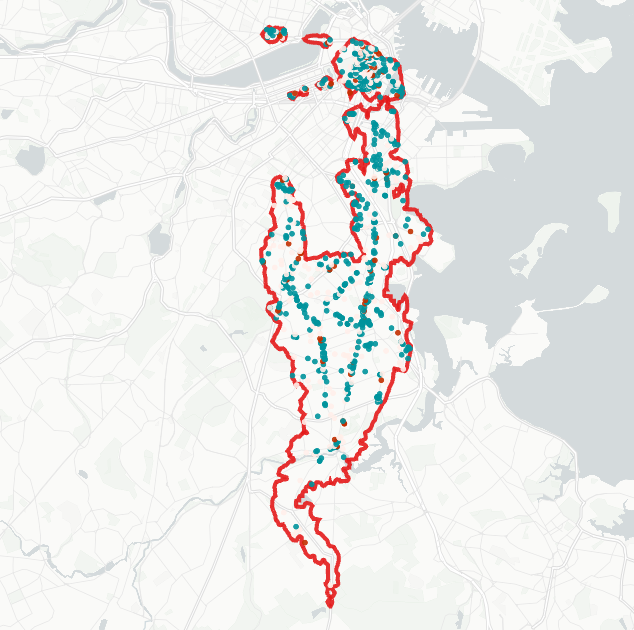}}
    \subfloat[]{\includegraphics[width=0.45\linewidth]{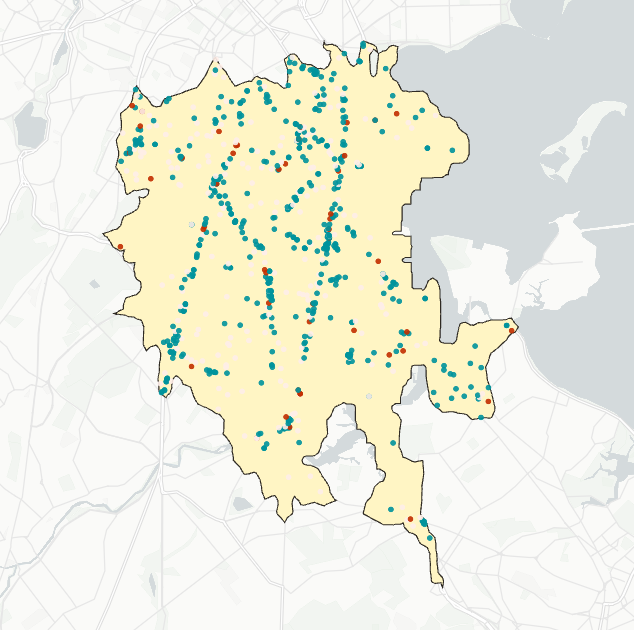}}
    \caption{Illustration of accessible services within the isochrone for each mode of transportation: (a) walking, (b) bicycle, (c) public transit, (d) driving.}
    \label{fig:services}
\end{figure}

We utilize a publicly available POI dataset to collect the numbers of services of various types within different regions. Each POI datum contains the geographic location of a specific service and its service type. By superimposing this data onto the isochrones, we can count the number of accessible services from a node $i\in\mathcal{V}$ within specific time frames. 
For instance, Fig. \ref{fig:neighborhoods} illustrates the location of services and neighborhoods in Boston, MA, USA. The (green) isochrone shows the accessible region within 30 minutes of driving, and the services within this isochrone cover most of the services. 
Figure \ref{fig:services} visualizes the distribution of essential and non-essential services within an isochrone for each mode of transportation. The points represent different service locations, with color variations indicating the types of services. Essential services, such as medical facilities and schools, are marked in distinct colors, emphasizing their significance in evaluating mobility equity.


\begin{remark}
    In our MI, we utilize isochrones that estimate the reachable distance within a time threshold considering average traffic conditions across the day on each road. However, the evaluation of mobility indices can also be conducted under different traffic conditions, e.g., a specific hour of the day, and different time thresholds, e.g., a smaller time threshold. The final evaluation of the MIs is sensitive to these choices, and different selections may yield different outcomes for any city.
\end{remark}

\subsection{Mobility Equity Metric}

While the MI offers an approach to quantify the accessibility and cost of mobility from any node of a transportation network, evaluating equity across the network should account for the fair distribution of the MI among the different nodes.
To this end, we employ a Gini coefficient \cite{gastwirth1972estimation} modified to account for population and define a mobility equity metric for the complete transportation network as follows.

\begin{definition} \label{def:mem}
    \vspace{3pt}
    Given MI $\varepsilon_i$ for each node $i\in\mathcal{V}$ of graph $\mathcal{G}$, the mobility equity metric (MEM) is 
    \begin{equation} \label{eq:mem}
        \mathrm{MEM}(\mathcal{G}) = 1-\frac{\sum_{i\in\mathcal{V}} \sum_{j\in\mathcal{V}} p_i p_j \left|\varepsilon_i - \varepsilon_j\right|}{2(\sum_{i\in\mathcal{V}} p_i)(\sum_{i\in\mathcal{V}} p_i \varepsilon_i)},
    \end{equation}
    where $p_i$ is the population of travelers at node $i \in \mathcal{V}$. 
\end{definition}
This metric provides a value in the range of $[0,1]$, with a value of $1$ indicating uniform MI across all regions. This is a direct consequence of using the Gini coefficient to evaluate the pairwise differences between MIs across the network in the second term in \eqref{eq:mem}, which is $0$ when $\varepsilon_i = \varepsilon_j$ for all $i, j \in \mathcal{V}$. In other words, if $\mathrm{MEM}(\mathcal{G})=1$, all regions exhibit the same level of accessibility to services after accounting for factors such as user cost, travel time, and transportation modes. For all $\mathrm{MEM}(\mathcal{G}) < 1$, the use of the Gini coefficient provides a way to measure the overall discrepancy in MIs of the populations residing at different nodes. 
To better account for these discrepancies, we incorporate the population at each node into the Gini coefficient in \eqref{eq:mem}. Doing so ensures that the equity is measured with respect to travelers rather than nodes in the network and mitigates biases emerging from uneven population distributions.

\begin{remark}
\vspace{3pt}
    In our prior work \cite{Bang2023mem}, MEM was defined as the average MI across regions. However, the revised MEM provided in Definition \ref{def:mem} is superior because (a) it takes values in $[0,1]$, allowing comparisons between different networks, (b) it is a true measure of equity that increases only if areas with low MI are improved, and (c) it includes the normalization of service counts to ensure that each service contributes fairly to the overall measure of equity.
\end{remark}

\begin{remark}
\vspace{3pt}
    In practice, when evaluating the MEM of a city's transportation network, we can only include finitely many nodes. However, the selection and omission of certain nodes can have a major impact on the eventual outcome of MEM evaluation. Thus, it is important to select a socially, economically, demographically, and geographically diverse set of nodes to represent the city's population.
\end{remark}

\begin{remark}
\vspace{3pt}
    When deciding how to partition the geography of a city into nodes for a transportation network, we consider one node per neighborhood or community, as defined by the U.S. census data. However, when deciding on allocating populations to nodes, it is crucial to explore different allocations by considering bigger or smaller spatial neighborhoods for which the assumption of uniform price sensitivity for all individuals within that area remains tenable. By ensuring that the purchasing power of all individuals at any node is comparable, we can ensure that the resulting MEM value is representative of the true equity within the city.
\end{remark}

\begin{table}[]
    \centering
    \caption{Priority level of services}
    \begin{tabular}{ |p{2cm}|p{0.75cm}||p{2.4cm}|p{0.75cm}|}
     \hline
     Service Type& $\beta$ &Service Type& $\beta$\\
     \hline
     Cafe   & 1/23    &Restaurant&  2/23\\
     Fitness&   2/23  &School& 3/23\\
     Hospital& 3/23 &Stadium&  1/23\\
     Market   & 3/23 &Theater&  1/23\\
     Park & 2/23 &Place of Worship&2/23\\
     Pharmacy& 3/23 &- &-\\
     \hline
    \end{tabular}
    \label{tab:priority}
\end{table}

\subsection{Mobility Equity Analysis for Major U.S. Cities}
\label{subsection:data analysis}


\begin{figure}[th]
    \centering
    \subfloat[]{\includegraphics[width=0.8\linewidth]{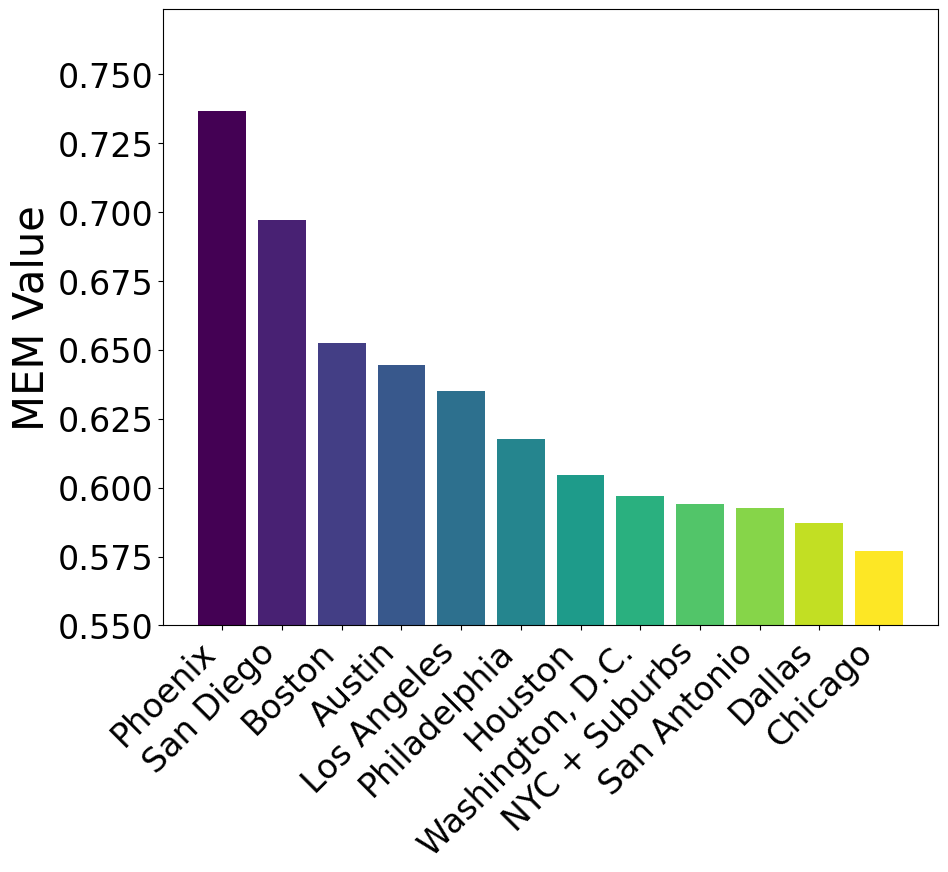} \label{fig:all cities MEM all s}}\\
    \subfloat[]{\includegraphics[width=0.8\linewidth]{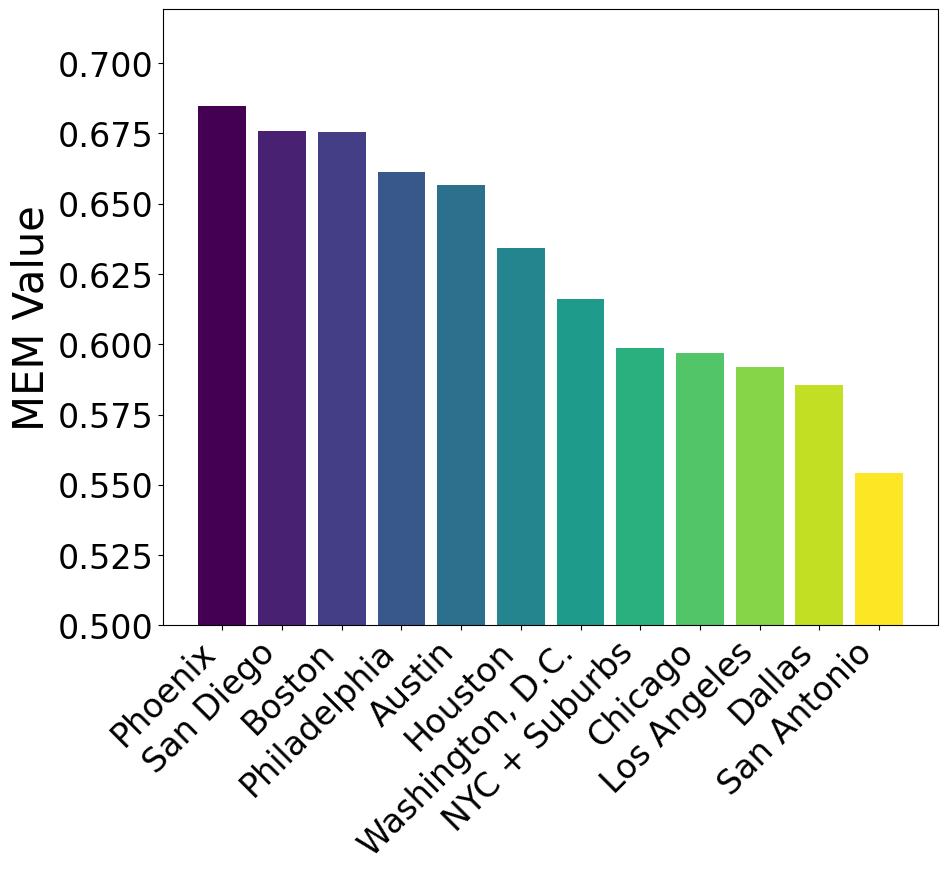} \label{fig:all cities MEM ess s}}
s    \caption{MEM values of 12 major cities in the United States: (a) considering all services and (b) only considering essential services}
    \label{fig:all cities MEM}
\end{figure}

\begin{figure*}[th]
    \centering
    \subfloat[]{\includegraphics[width=0.28\linewidth]{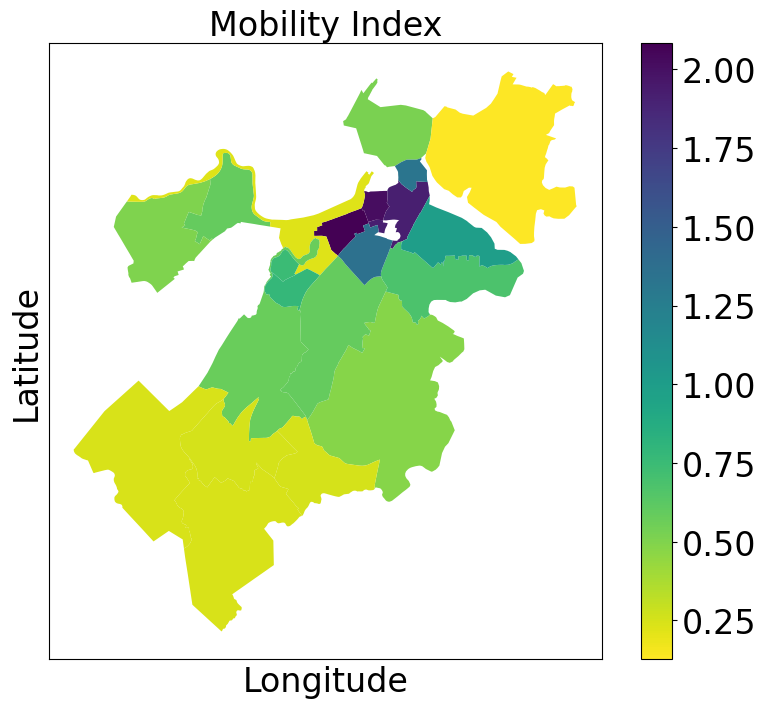}}
    \subfloat[]{\includegraphics[width=0.28\linewidth]{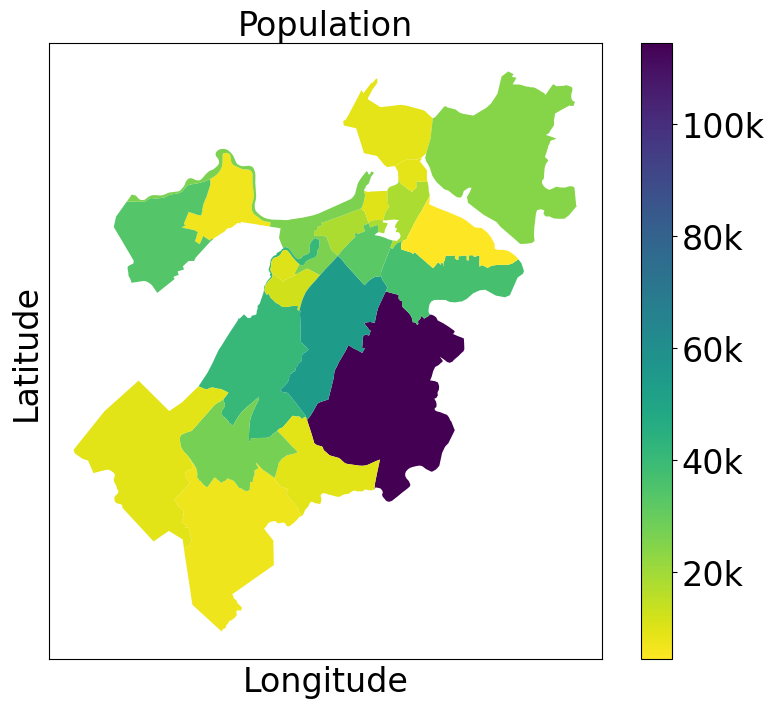}}
    \subfloat[]{\includegraphics[width=0.28\linewidth]{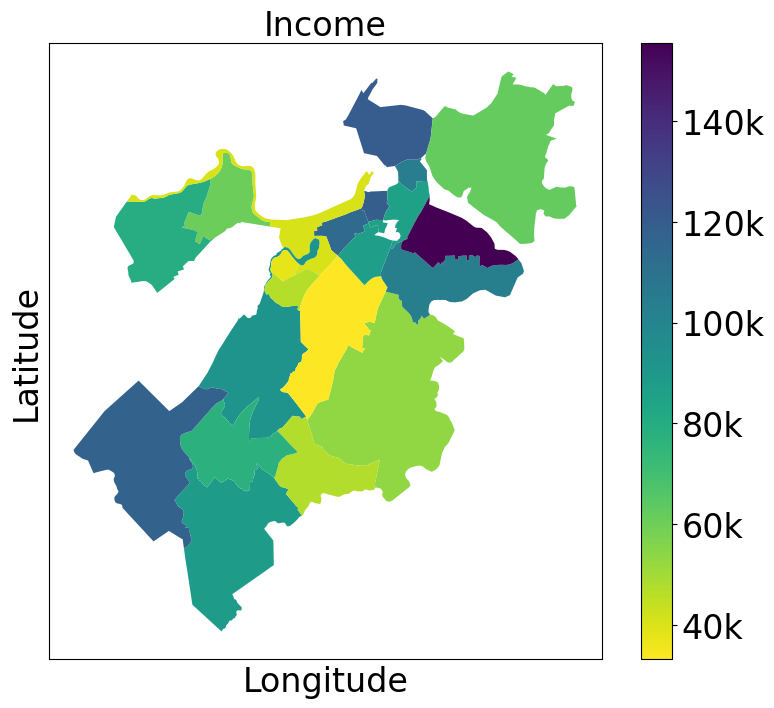}}
    \caption{Heatmaps in Boston city: average (a) mobility index, (b) population, (c) income.}
    \label{fig:heatmap_boston}
\end{figure*}

\begin{figure*}[th]
    \centering
    \subfloat[]{\includegraphics[width=0.285\linewidth]{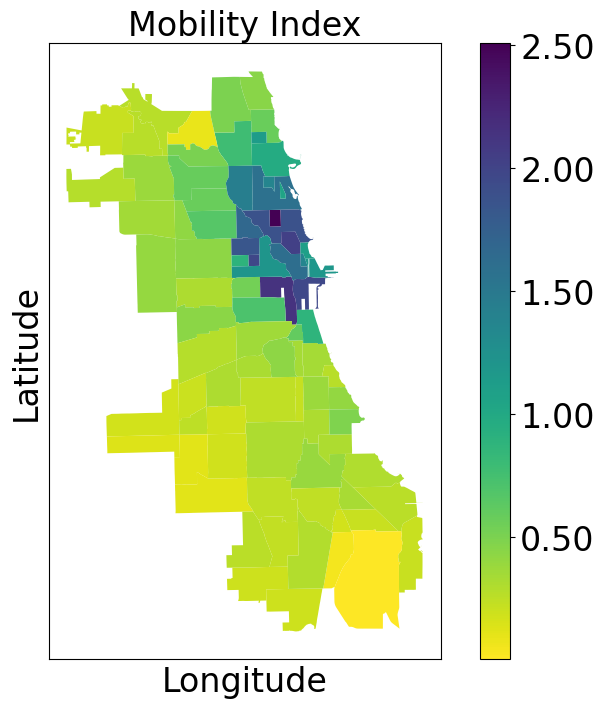}}
    \subfloat[]{\includegraphics[width=0.28\linewidth]{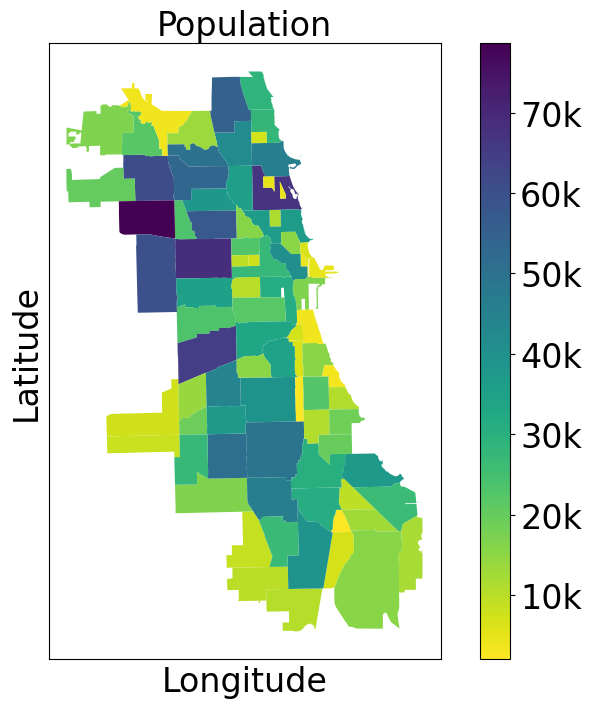}}
    \subfloat[]{\includegraphics[width=0.29\linewidth]{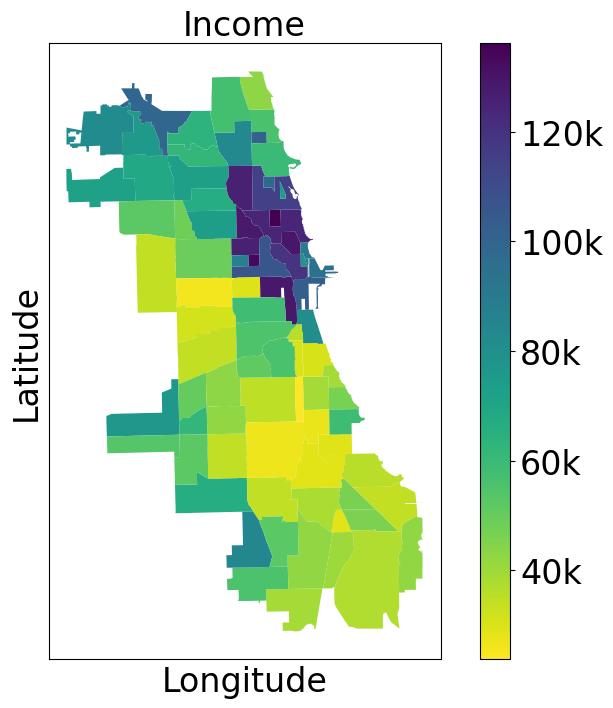}}
    \caption{Heatmaps in Chicago: average (a) mobility index, (b) population, (c) income.}
    \label{fig:heatmap_chicago}
\end{figure*}

\begin{figure*}[th]
    \centering
    \subfloat[]{\includegraphics[width=0.29\linewidth]{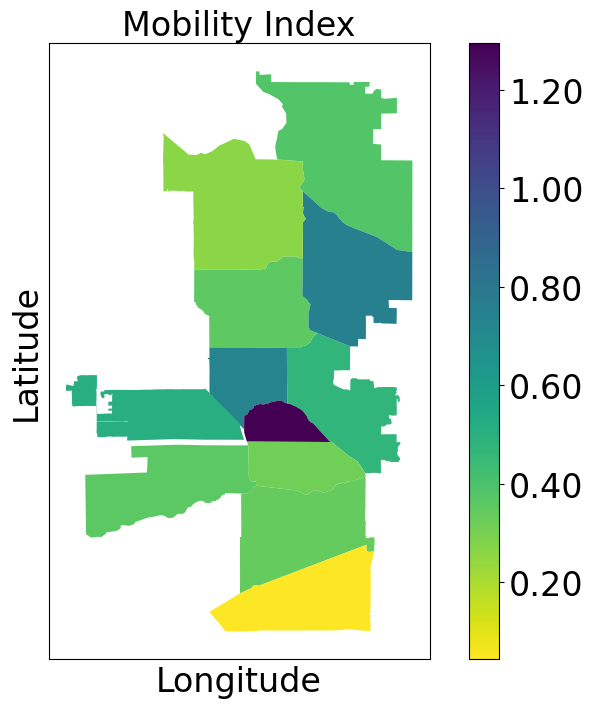}}
    \subfloat[]{\includegraphics[width=0.28\linewidth]{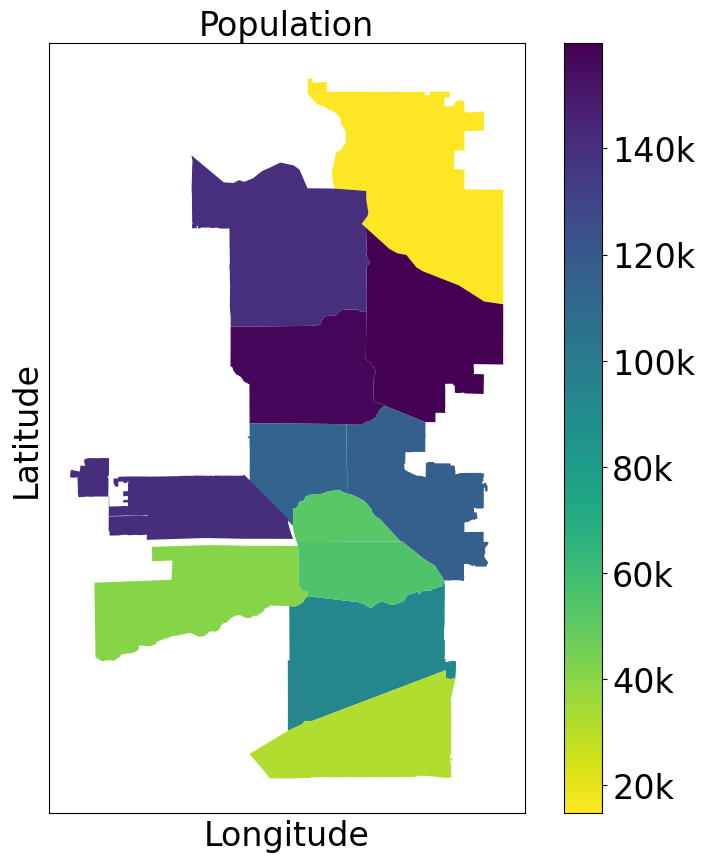}}
    \subfloat[]{\includegraphics[width=0.28\linewidth]{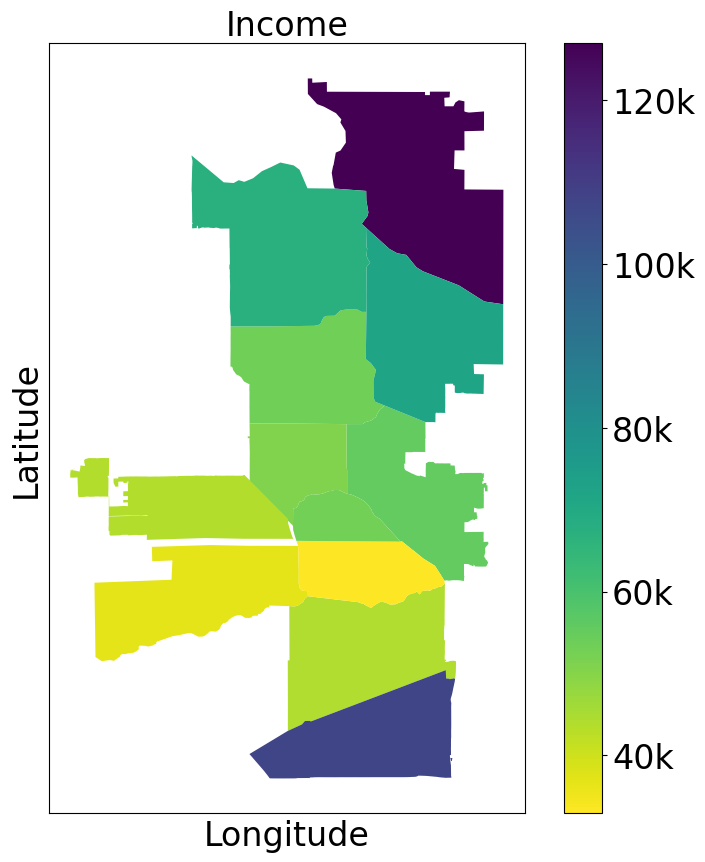}}
    \caption{Heatmaps in Phoenix: average (a) mobility index, (b) population, (c) income.}
    \label{fig:heatmap_phoenix}
\end{figure*}

\begin{figure*}[th]
    \centering
\end{figure*}

\begin{figure*}[th]
    \centering
    \subfloat[]{\includegraphics[width=0.29\linewidth]{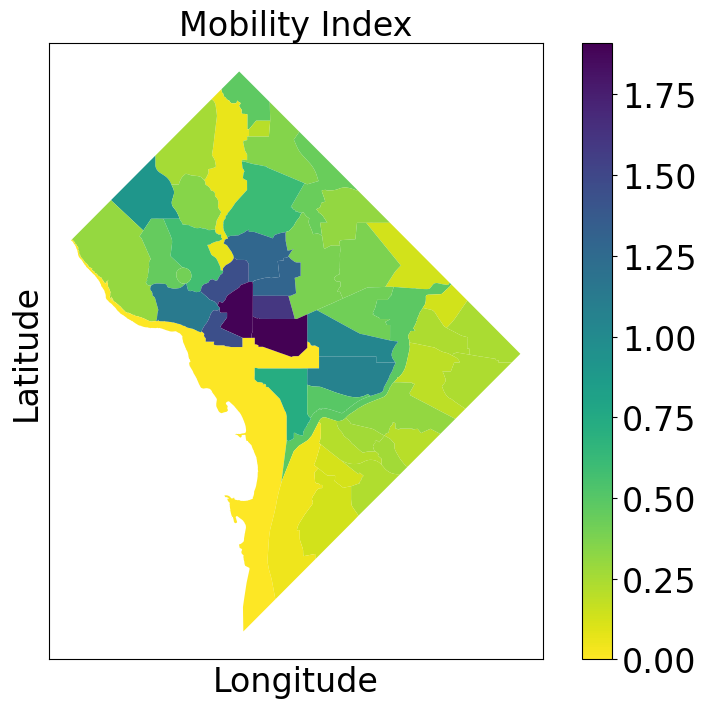}}
    \subfloat[]{\includegraphics[width=0.28\linewidth]{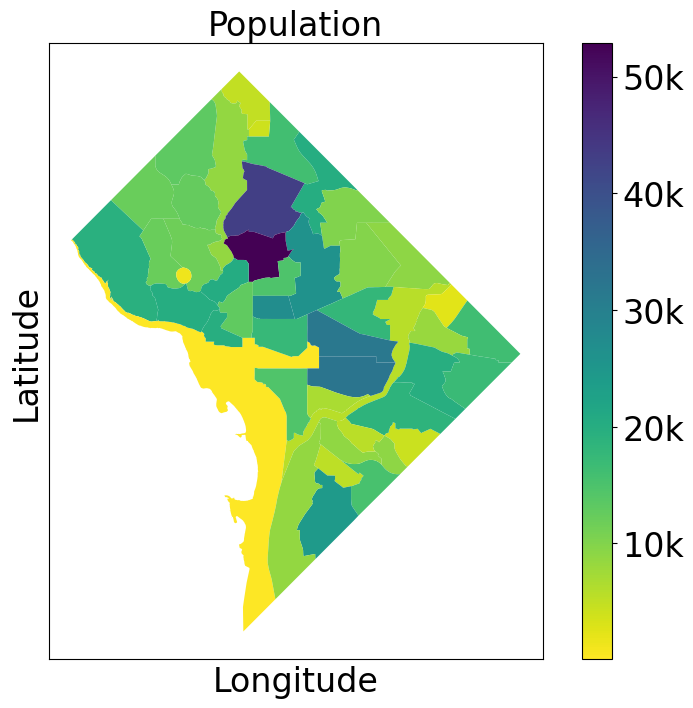}}
    \subfloat[]{\includegraphics[width=0.29\linewidth]{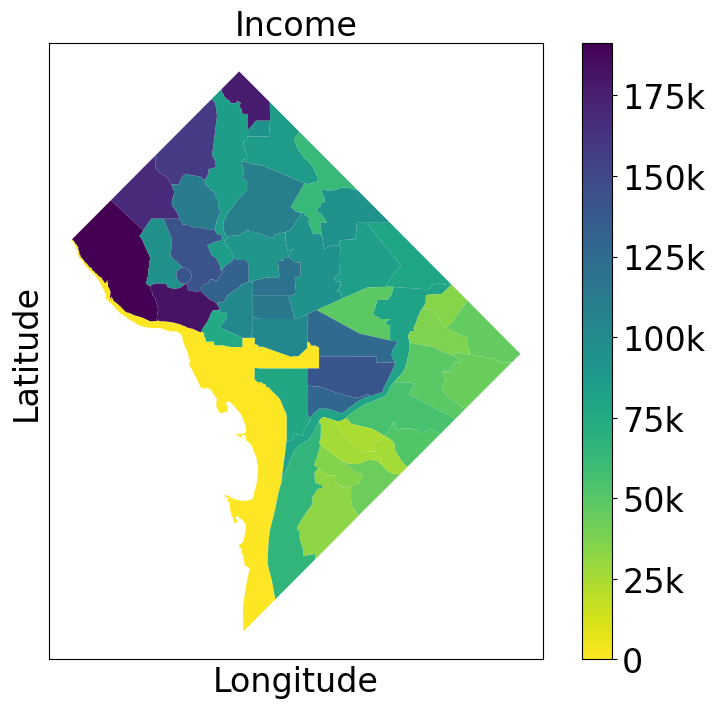}}
    \caption{Heatmaps in Washington D.C.: average (a) mobility index, (b) population, (c) income.}
    \label{fig:heatmap_dc}
\end{figure*}

In this subsection, we demonstrate the applicability and insights of our proposed MEM through a detailed analysis of mobility equity across $12$ major cities in the United States.
We begin by dividing each city into corresponding communities as found in census tract data published by the cities \cite{Census}.
Leveraging census tract data and demographic information, we evaluate the MI for each community, incorporating accessibility to eleven essential services listed in Table \ref{tab:priority} from the centroid of the community's geographical region and the average income level of the community's residents.
To enumerate the services across each city, we collect POI data using the TOMTOM API \cite{TOMTOM} to mark the exact geographical position of each service in and around the city. Then, starting from each community's centroid, we use the Isoline API from Geoapify \cite{Geoapify} to generate isochrones corresponding to four modes of transportation, namely, walking, bicycles, public transit, and passenger cars under approximated traffic conditions. 
We select the cost $c_m$ for each mode $m$ based on the monetary cost required to maintain/operate the mode. For example, we utilize the ticket fare for the cost of public transportation, consider the total price of a bicycle divided by the overall usage as the cost of using the bicycle for a single trip, and assume $\$0$ for walking. For private vehicles, we infer the total cost per passenger mile using travel reimbursement in the city and multiply it by the average travel distance within the $30$-minute time threshold.
When evaluating the accessibility, the priority level $\beta^s$ of each service type $s \in \mathcal{S}$ is selected from Table \ref{tab:priority}, considering a distinction between essential services, convenience services, and leisure services.
Similarly, to correctly capture the trade-off between the cost of traveling using a specific mode of transportation and the accessibility, we determine price sensitivity $\kappa_i$ by the relative income level of a neighborhood $i$ to the maximum income across the city. By incorporating the maximum income within the city, we normalize the price sensitivity with respect to the cost of living in the city. 

In Fig. \ref{fig:all cities MEM all s}, we present the results of our analysis for $12$ cities, ranked according to their MEM values after incorporating the services in Table \ref{tab:priority}. This figure reveals a range of MEM values across the cities, ranging from $0.577$ for Chicago. IL to $0.737$ for Phoenix, AZ. In Fig. \ref{fig:all cities MEM ess s}, we contrast the above MEM values against MEM values when considering only the essential services: hospitals, pharmacies, schools, and markets. We observe that essential MEM values showcase a smaller range, from $0.554$ in San Antonio, TX, and $0.685$ in Phoenix, AZ. Certain cities, e.g., Chicago, IL, and Philadelphia, PA, show an improvement in essential MEM compared to general MEM values. This can be attributed to these cities having a better distribution of essential services and possibly having a greater concentration of non-essential services, such as restaurants in the downtown region. In contrast, certain cities, e.g., Los Angeles, CA, show a decrease in essential MEM compared to regular MEM. Once again, this can be attributed to Los Angeles having a more even distribution of non-essential services and a concentration of essential services to a few regions. Ultimately, comparing the MEM values for both collections of services yields a surprising result: cities such as Chicago, which has a great reputation for public transit, may not have the highest mobility equity. 

\begin{figure*}
    \centering
    \subfloat[]{\includegraphics[width=0.4\linewidth]{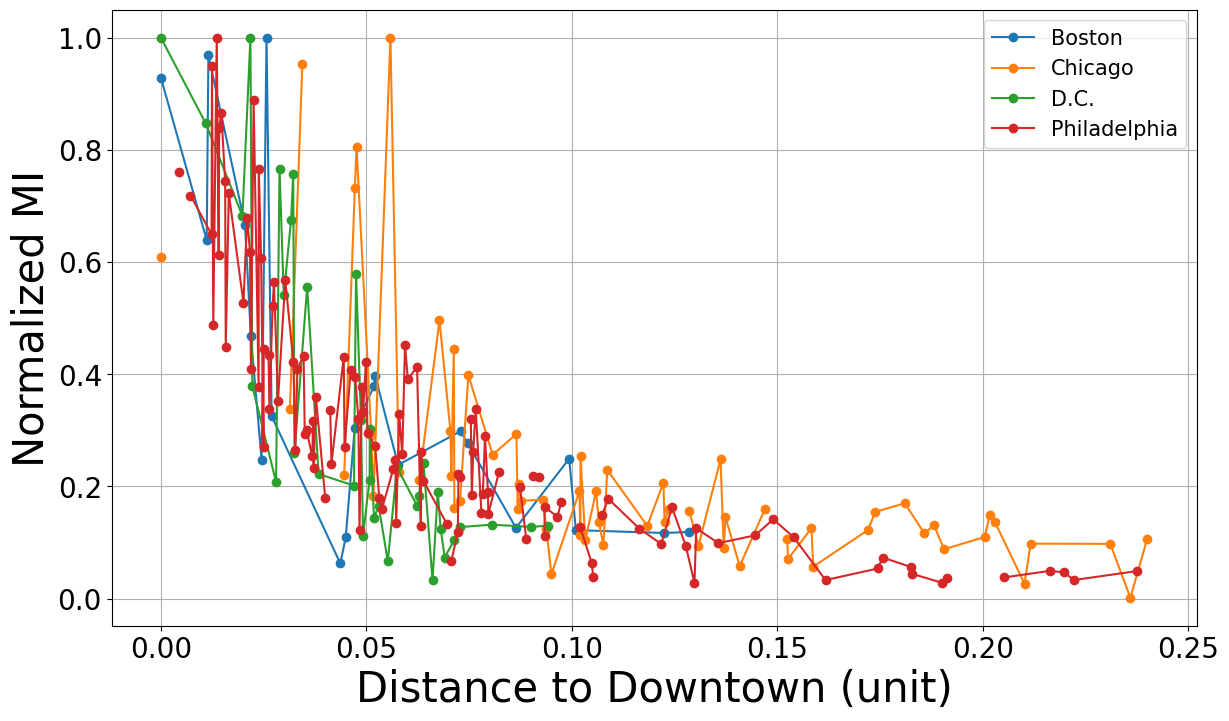}}
    \subfloat[]{\includegraphics[width=0.4\linewidth]{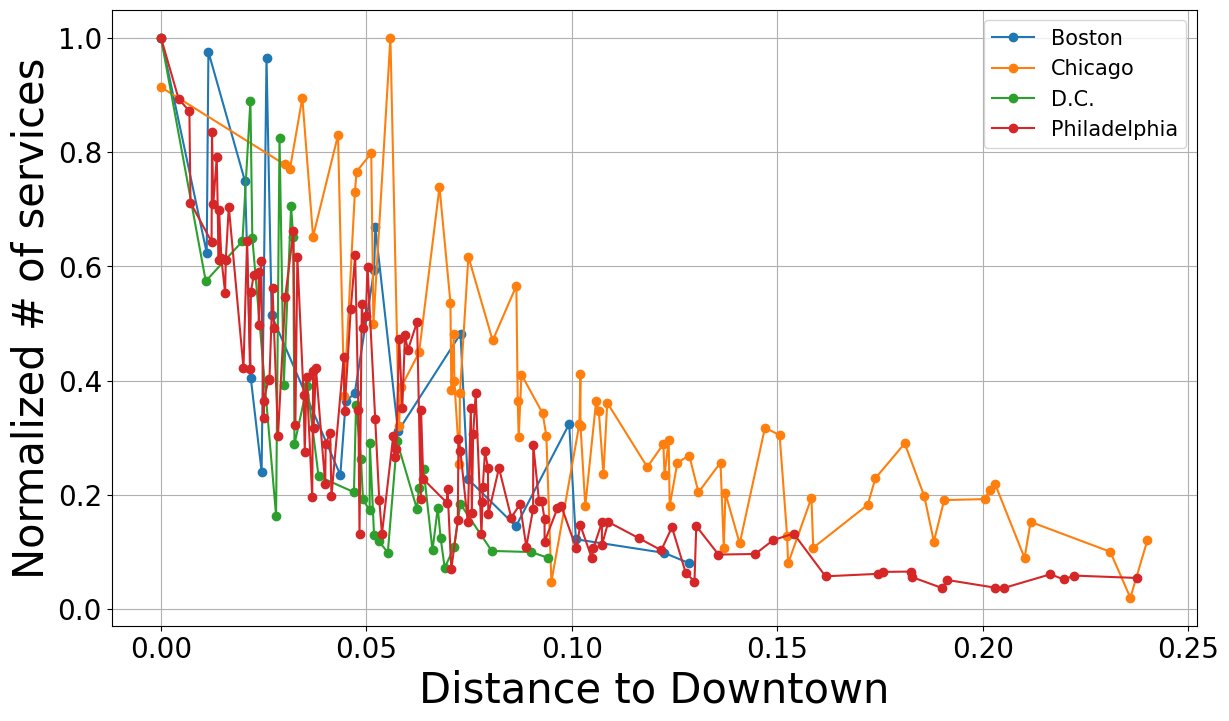}}\\
    \subfloat[]{\includegraphics[width=0.4\linewidth]{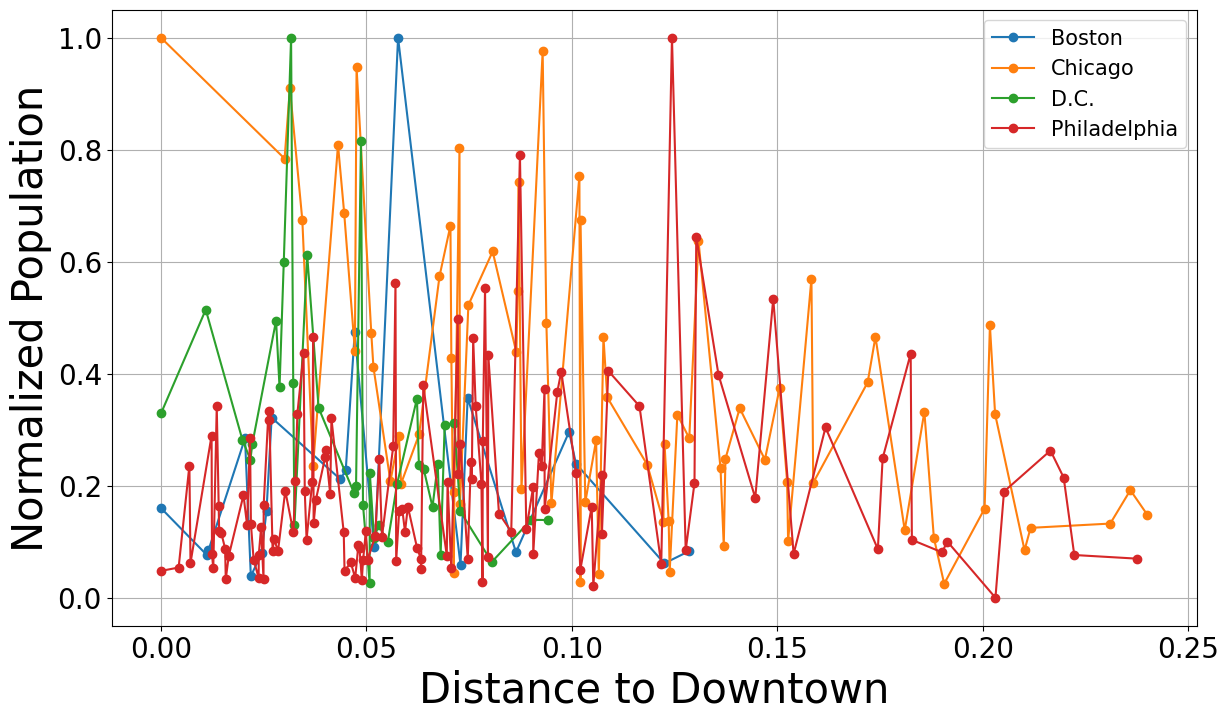}}
    \subfloat[]{\includegraphics[width=0.4\linewidth]{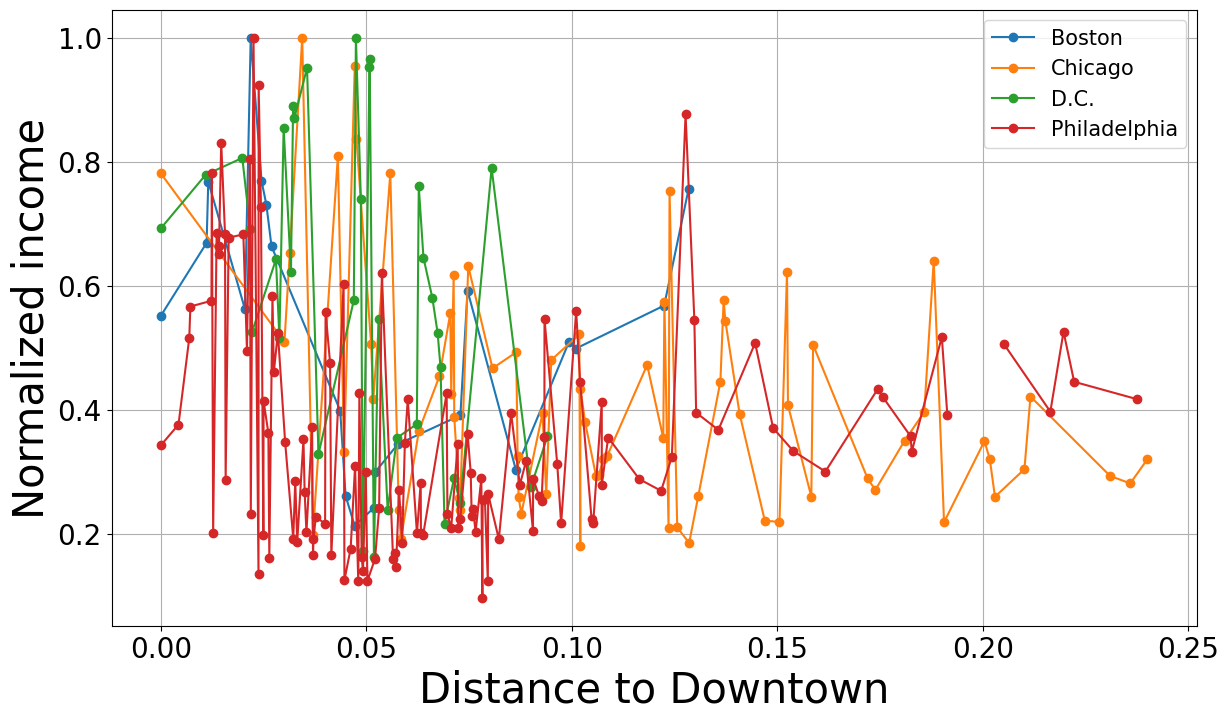}}
    \caption{Relationship between (a) MI, (b) the number of services, (c) populations, and (d) income relative to the distance from the downtown area.}
    \label{fig:distance_trend}
\end{figure*}

To better understand the factors that determine these eventual MEM values, we present heatmaps for MIs, income levels, and population counts across communities in four cities: Boston, Chicago, Phoenix, and Washington D.C. (see Figs. \ref{fig:heatmap_boston}-\ref{fig:heatmap_dc}). These figures reveal a consistent pattern across all four cities, with the highest MI values concentrated in the downtown areas and gradually decreasing as the distance from the city center increases.
This trend highlights the centralized nature of urban planning, where essential services and transportation infrastructure are often clustered in the heart of the city, leading to lower MI values in peripheral neighborhoods.
Figure \ref{fig:distance_trend} delves deeper into the relationship between MI and the distance from the downtown area.
Specifically, it demonstrates a clear negative correlation between MI and distance, with neighborhoods farther from the city center experiencing a sharp decline in accessible services.
Meanwhile, it also shows that the population does not have a clear correlation with distance and that high-income groups reside relatively closer to the Downtown area. This observation suggests that affluent communities often benefit from a balance between accessibility and livability, residing in areas that offer convenient access to essential services without the congestion and density of the central business district.
Considering the individual heatmaps explains the high MEM value of Phoenix against the low MEM value of Chicago. From Fig. \ref{fig:heatmap_phoenix}, we observe that the MI is evenly distributed across most of the higher population regions of the city. Furthermore, the low-income areas of Phoenix are closer to the center of the city and, thus, have easier access to most services and high MI neighborhoods. In contrast, Fig. \ref{fig:heatmap_chicago} reveals that Chicago displays a very strong concentration of high MIs in the downtown region. However, a large population of Chicago's low-income residents reside in low MI regions far away from the city's downtown area. This disparity in service access contributes significantly to the city's low MEM value.


\begin{figure}
    \centering
    \subfloat[]{\includegraphics[width=0.9\linewidth]{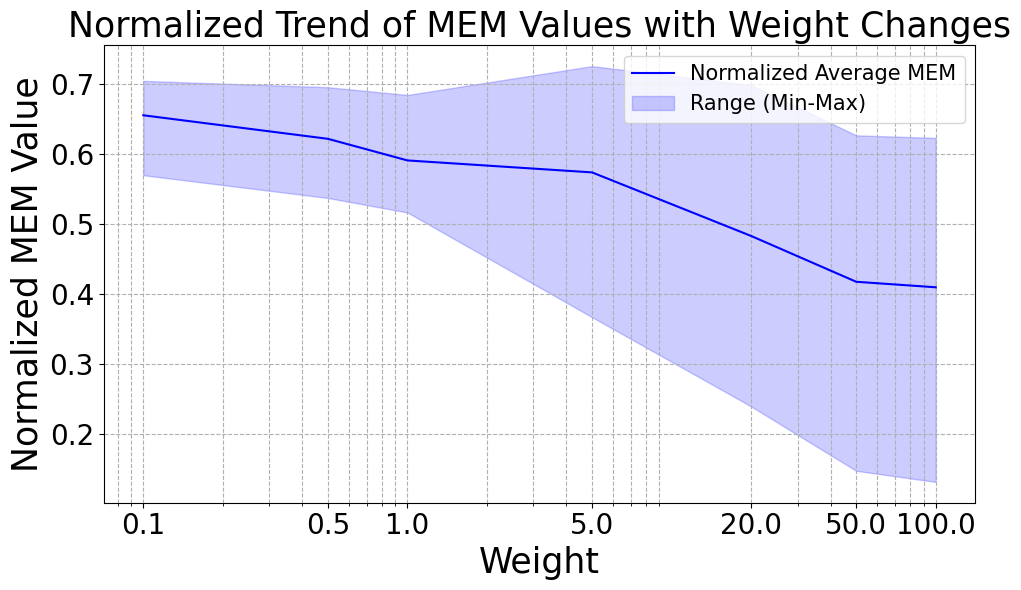}}\label{fig:mem_trend}\\  
    \subfloat[]{\includegraphics[width=0.9\linewidth]{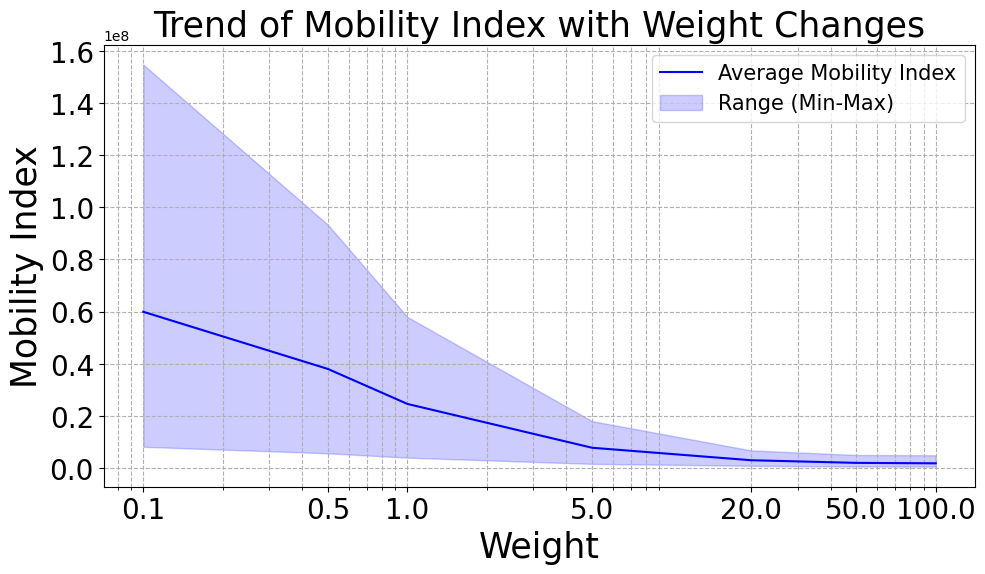}}\label{fig:mi_trend}
    \caption{Impact of price sensitivity $\kappa$ on (a) MEM and (b) MI in the city of Boston.}
    \label{fig:weight_trend}
\end{figure}

Next, we explore the impact of the price sensitivity hyperparameter on MEM and MI values in the city of Boston in Fig. \ref{fig:weight_trend}. In these graphs, the weight refers to a factor we use to scale the price sensitivity $\kappa_i$ for all $i \in \mathcal{V}$. This change in the price sensitivity can be understood as a change in the overall cost of living within the city, i.e., an increase in the cost of living is modeled by an increase in $\kappa_i$. Since this increase impacts the price sensitivity of all neighborhoods, it leads to a decrease in the average MI values across the city. Furthermore, this increase in the cost of living may disproportionately affect people from different demographics by limiting the modes of transportation feasible for them. Thus, it also leads to decreased MEM values across the city.
This finding emphasizes the importance of considering the affordability of transportation options when assessing mobility equity. High costs can disproportionately affect lower-income communities and exacerbate existing inequities in accessibility. The choice of the hyperparameter $\kappa_i$ for each $i \in \mathcal{V}$ is essential to correctly assessing the equity in a city.

Finally, we developed an interactive demonstration tool (\url{https://horatioj.shinyapps.io/MobDashboard}). This tool provides comprehensive visualizations, including heatmaps and MEM values for various time thresholds, allowing urban planners to explore the nuances of mobility equity across different temporal and spatial scales.

\section{Routing for Emerging Transportation Systems} \label{sec:routing}

Given that the proposed MEM quantifies transportation equity, in this section, we explore how to optimize equity within an intelligent transportation network using CAVs. 
To achieve this goal, we present a routing framework that incorporates the MEM as an upper-level objective for a system planner that makes routing suggestions to vehicles corresponding to all modes of transit in the network. 
In our transportation network, we consider public transit consisting of CAVs and private (either automated or human-driven) vehicles.
Public transit CAVs always adhere to the routes suggested by a system planner, while private vehicles may or may not follow the recommendations.
Therefore, we aim to develop MEM-optimizing route suggestions for compliant vehicles while accounting for the choices made by non-compliant vehicles.

As in Section \ref{sec:mem}, the transportation network is modeled as a directed graph $\mathcal{G} = (\mathcal{V},\mathcal{E})$, with a set of modes $\mathcal{M}$ and service types $\mathcal{S}$. In particular, we consider two primary modes in the network: public transit and private vehicles.
The set $\mathcal{N}=\{1,\dots,N\}$, $N\in\mathbb{N},$ represents a collection of possible trips, each defined by an origin-destination pair.
Origins represent travel demand points, while destinations are locations that offer various services.
Each trip $n \in \mathcal{N}$ is associated with the origin $o_n\in\mathcal{O} \subseteq \mathcal{V}$ and destination $d_n\in\mathcal{D} \subseteq \mathcal{V}$, and the corresponding travel demand rate for trip $n \in \mathcal{N}$ using mode $m \in \mathcal{M}$ is denoted by $\alpha_{m,n}\in\mathbb{R}_{>0}$.

Next, We formulate the vehicle-flow optimization problem. We define $x^{ij}_{m,n} \in \mathbb{R}_{\geq0}$ as the flow of compliant vehicles on edge $(i,j) \in \mathcal{E}$ for trip $n$ using mode $m$.
The total complying-vehicle flow on edge $(i,j)$ is calculated as $x^{ij}=\sum_m\sum_n h_m \cdot x^{ij}_{m,n}$, where $h_m$ represents the road occupancy of mode $m$.
For instance, we set public transportation occupancy $h_\mathrm{public}$ to be $0.8$ as multiple passengers use a single vehicle, whereas private vehicle occupancy $h_\mathrm{private}$ is $1$.
The flow of non-compliant vehicles on edge $(i,j)$ is denoted by $q^{ij} \in \mathbb{R}_{\geq0}$.

Given the cumulative flows of both compliant and non-compliant vehicles on edge $(i,j)$, we utilize a travel time function proposed by \textit{Bureau of Public Roads (BPR)}, which is also known as BPR function:
\begin{equation}
    t^{ij}(x^{ij}+q^{ij}) = t^{ij}_0 \cdot \left( 1+0.15 \left(\frac{x^{ij}+q^{ij}}{\gamma^{ij}}\right)^4\right),
\end{equation}
where $t^{ij}_0 \in \mathbb{R}_{\geq0}$ is the free-flow travel time and $\gamma^{ij} \in \mathbb{R}_{\geq0}$ is the road capacity on edge $(i,j)$.


Next, we introduce two routing problems: the first aims to find the system-wide optimal solution for compliant vehicles, while the second models a selfish routing mechanism followed by non-compliant vehicles.

\textit{1) System-centric routing} provides compliant vehicles with the system-wide minimum time routes by solving the following optimization problem from the system planner.
\begin{problem}[System-Centric Routing] \label{pb:system-centric}
    \begin{equation}
        \begin{aligned}
            &\minimize_{\{x^{ij}_{m,n}\}} ~ \sum_{m\in\mathcal{M}} w_m \left\{ \sum_{n\in\mathcal{N}}\sum_{(i,j)\in\mathcal{E}} t^{ij}(x^{ij}+q^{ij})\cdot x^{ij}_{m,n} \right\}\\
            \mathrm{s.t.} &~ \sum_{k:(j,k)\in\mathcal{E}} x^{jk}_{m,n} = \alpha_{m,n},~~\forall m\in\mathcal{M},n\in\mathcal{N},j=o_n,\\
            & \sum_{i:(i,j)\in\mathcal{E}} x^{ij}_{m,n} = \alpha_{m,n},~~\forall m\in\mathcal{M},n\in\mathcal{N},j=d_n,\\
            & \sum_{i:(i,j)\in\mathcal{E}} x^{ij}_{m,n} = \sum_{k:(j,k)\in\mathcal{E}} x^{jk}_{m,n},~~\forall m\in\mathcal{M},n\in\mathcal{N},j\in\bar{\mathcal{V}},
        \end{aligned}
    \end{equation}
    where $w_m$ is a weight on mode $m$, $\bar{\mathcal{V}} = \mathcal{V}\setminus\{o_n,d_n\}$ for all $o_n\in\mathcal{O}$, $d_n\in\mathcal{D}$.
\end{problem}
The constraints ensure the flow aligns with the demand rate and connects corresponding origins and destinations.
Problem \ref{pb:system-centric} is a convex optimization problem since the objective function is convex in its domain and the inequality constraints are linear.
Note that the solution to Problem \ref{pb:system-centric} depends on the weights $w_m$ for each mode $m \in \mathcal{M}$, which implies that MEM also depends on the choice of the weights $w_m$. Thus, we present later a problem of selecting proper weights to acquire the maximum MEM.

\textit{2) Non-compliant vehicle routing} captures the bounded rationality in the decision-making of non-compliant private vehicles.
A common approach to predicting the trends of the non-complaint vehicle flow is to utilize the concept of \textit{Wardrop equilibrium} \cite{wardrop1952road}. This concept is similar to Nash equilibrium with multiple players, where all players (in this case, non-compliant vehicles) attempt to achieve their best performance at any given time.
Although Wardrop equilibrium explains non-compliant vehicle flow behavior, it assumes that all vehicles make perfectly strategic decisions. Moreover, to make such decisions, these vehicles must be able to anticipate other vehicles' movements with perfect information or based on previous experience.
These underlying assumptions may not hold in real-world scenarios, especially when travelers' demands vary daily.
To capture this imperfect decision-making, we employ a cognitive hierarchy model \cite{Bang2023mem} for non-compliant vehicles' behavior.
This model introduces different levels of cognition among human drivers, where a driver with a higher cognition level can anticipate the decisions of other drivers with lower cognition levels.

For each trip $n \in \mathcal{N}$ and $\ell$-level non-compliant vehicle, $\ell=0,1,\dots,L$, we let $q_{\ell,n}$ be the demand rate from the origin $o_n \in \mathcal{O}$ to the corresponding destination $d_n \in \mathcal{D}$.
For an $\ell$-level non-compliant vehicle traveling for trip $n$, we define assignment vector $A_{\ell,n}\in 2^{|\mathcal{E}|}$ where the element $a^{ij}_{\ell,n}$ takes value of $1$ if the $\ell$-level non-compliant vehicle for trip $n$ uses the edge $(i,j)$ and takes value of $0$ otherwise. Then, we solve the following problem to predict the routes of all the non-compliant vehicles.

\begin{problem}[Non-compliant Vehicle Routing] \label{pb:selfish}
\vspace{3pt}
Each driver with cognition level $\ell\in\{0,1,\dots,L\}$ selects the shortest-time path based on the anticipated flow of lower-level drivers, i.e.,
\begin{equation}
        \begin{aligned}
        \minimize_{\{a_{\ell,n}^{ij}\}}~& \sum_{n\in\mathcal{N}}\sum_{(i,j)\in\mathcal{E}} t^{ij}\left(x^{ij}+\sum_{l=0}^{\ell-1}q_{l}^{ij}\right)\cdot a_{\ell,n}^{ij} \\
        \mathrm{s.t.~} & \sum_{k:(j,k)\in\mathcal{E}} a^{jk}_{\ell,n} = 1,~\forall n\in\mathcal{N},j=o_n,\\
        & \sum_{i:(i,j)\in\mathcal{E}} a^{ij}_{\ell,n} = 1,~\forall n\in\mathcal{N},j=d_n,\\
        & \sum_{i:(i,j)\in\mathcal{E}} a^{ij}_{\ell,n} = \sum_{j:(j,k)\in\mathcal{E}} a^{jk}_{\ell,n},\\
        &\hspace{13ex}\forall n\in\mathcal{N},j\in\mathcal{V}\setminus\{o_n,d_n\},
        \end{aligned}
    \end{equation}
    where $q_\ell^{ij} = \sum_{n\in\mathcal{N}}q_{\ell,n} \cdot a_{\ell,n}^{ij}$ and $q_{\ell,n}$ represents the flow of $\ell$-level non-compliant vehicles for trip $n$.
\end{problem}
Note that $q^{ij}$ is the total flow of non-compliant vehicles on edge $(i,j)$ and $x^{ij}$ in Problems \ref{pb:system-centric} and \ref{pb:selfish} is the compliant-vehicle flow. We assume that all vehicles can anticipate the flows induced by compliant vehicles through knowledge of public transit schedules, typical traffic situations at certain hours, etc.

Problem \ref{pb:selfish} is a binary optimization, which we solve using Dijkstra algorithm.
Solution to Problem \ref{pb:selfish} is the shortest-time path for a single non-compliant vehicle considering the compliant vehicles' flow and the lower-level vehicles. Then, we consider all the non-compliant vehicles at the same cognition level for a given origin-destination pair to take the same route since they do not know the existence of the other vehicles at the same level and their impact on the actual flow.

Solutions to Problem \ref{pb:system-centric} and Problem \ref{pb:selfish} constitute the net flow on the network. This allows us to estimate the travel time $t_m^{o,d}$ for each trip $(o_n,d_n)$, for $n\in\mathcal{N}$, and each mode $m\in\mathcal{M}$.
Consequently, we can evaluate MI at each origin $o_n$ by counting the number of accessible services using mode $m$ within the time threshold $\tau_m$, i.e., $\sigma_{o_n,m}^s(\tau_m) = \sum_{d\in\mathcal{D}} \eta^s_d\cdot\mathbb{I}\Big[t_m^{o_n,d} \leq \tau_m\Big]$, where $\eta^s_d$ is the number of services with type $s$ at destination $d$ and $\mathbb{I}$ is an indicator function.
Note that the MI resulted from solutions to Problems \ref{pb:system-centric} and \ref{pb:selfish} depends on the weight $w_m$, $m\in\mathcal{M}$.
Thus, we can formulate an optimization problem to obtain better MEM with respect to the weights.


\begin{problem}[Mobility Equity Maximization] \label{pb:mem}
\vspace{3pt}
\begin{equation}
    \begin{aligned}
        \maximize_{w}&~~ \textit{MEM}\\
        \mathrm{subject~to:}&~~ \delta^\mathrm{pv}(w) \leq \lambda,
    \end{aligned}
\end{equation}
where $\delta^\textrm{pv}$ is the average travel-time difference between compliant private vehicles and non-compliant vehicles, and $\lambda$ is the upper limit of the difference.
\end{problem}

We impose the upper limit on the travel-time difference to ensure that the suggested route does not force compliant vehicles to sacrifice their travel time over the limit $\lambda$.

\section{Simulation and Analysis} \label{sec:simulation}

\subsection{Numerical Simulation Setup}

To validate the effectiveness of our proposed routing framework in enhancing mobility equity, we conducted numerical simulations using a network representation of the Boston Metropolitan Area. The network, depicted in Fig. \ref{fig:Boston_Network}, consists of 45 intersections connecting 8 origins to 5 destinations.
We selected origins based on demographic data from specific Boston neighborhoods, considering their coordinates and population \cite{boston}. Similarly, the destinations represent locations of essential services, with multiple services at each node to reflect the real-world clustering of facilities where various stores and services are often within walking distance.

To closely mimic real-world conditions, we incorporated accurate road lengths and capacities using PTV Vissim software with Bing Maps. This information, combined with speed limits, allowed us to determine the capacity and free-flow travel time of each link.
The simulations considered two primary modes of transportation: private vehicles and public transportation. The cost per passenger mile for public transportation was set at $10\%$ of the cost for private vehicles to reflect the typical cost difference between these modes. To alleviate the computational burden of solving Problems \ref{pb:system-centric} and \ref{pb:selfish} multiple times, we used a piece-wise linear approximation of the BPR function as described in \cite{wollenstein2021routing}.

For evaluating MEM, it is required to calculate the average number of services accessible within the time threshold $\tau_{m}$ associated with each mode of transportation. To evaluate this quantity with the information on vehicle flows, we use the following formula: 
\begin{equation}
\sigma^s_{i,m}(\tau_m)=\sum_{d\in\mathcal{D}_s} \eta^s_d \cdot\mathbb{I}\;[t_m^{i,d}\leq \tau_m]
\end{equation}
where $t_m^{i,d}$ represents the average travel time from node $i$ to destination $d$. Note that the summation is designed to cover all different destinations included in $\mathcal{D}_s$, which includes all the destinations that are associated with the service $s$. For the indicator function, as discussed in \cite{Bang2023mem}, we use a continuous approximation to capture small deviations in travel time. The approximation of the indicator function is the following: 

\begin{figure}[th]
    \centering
    \includegraphics[width=0.9\linewidth]{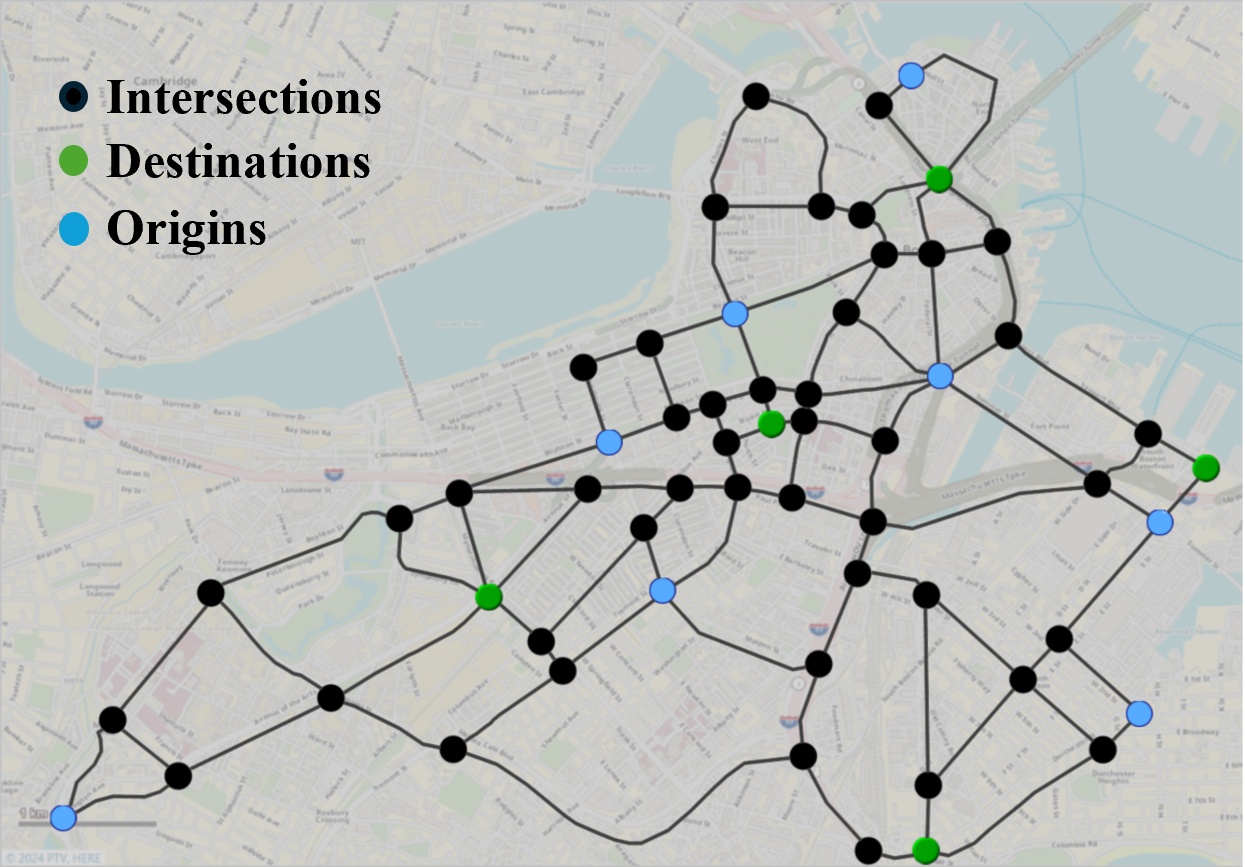}
    \caption{Road network in Boston.}
    \label{fig:Boston_Network}
\end{figure}

\begin{equation}
    \mathbb{I}\;[t\leq \tau_m] \simeq 1-\frac{1}{1+e^{-k(t-\tau_m)}} \label{eqn:approx}
\end{equation}
where $k\in \mathbb{R^{+}}$ is a parameter that allows to manipulate the slope of the indicator function appropriately according to how sensitive we want to be in small deviations in travel time.

\begin{remark}
    In fact, the method of approximating accessible services in \eqref{eqn:approx} is the generalization of gravity-based and isochronic measures of accessibility. With larger $k$, the approximation in \eqref{eqn:approx} converges to the actual indicator function $\mathbb{I}$ and yields the isochronic measure. On the other hand, smaller $k$ induces the number of accessible services $\sigma_{i,m}^s$ to smoothly decrease as travel time increases, which is identical to the gravity-based measure.
\end{remark}

\begin{figure*}[h!]
    \centering
    \begin{subfigure}[b]{0.49\textwidth}
        \includegraphics[width=\textwidth]{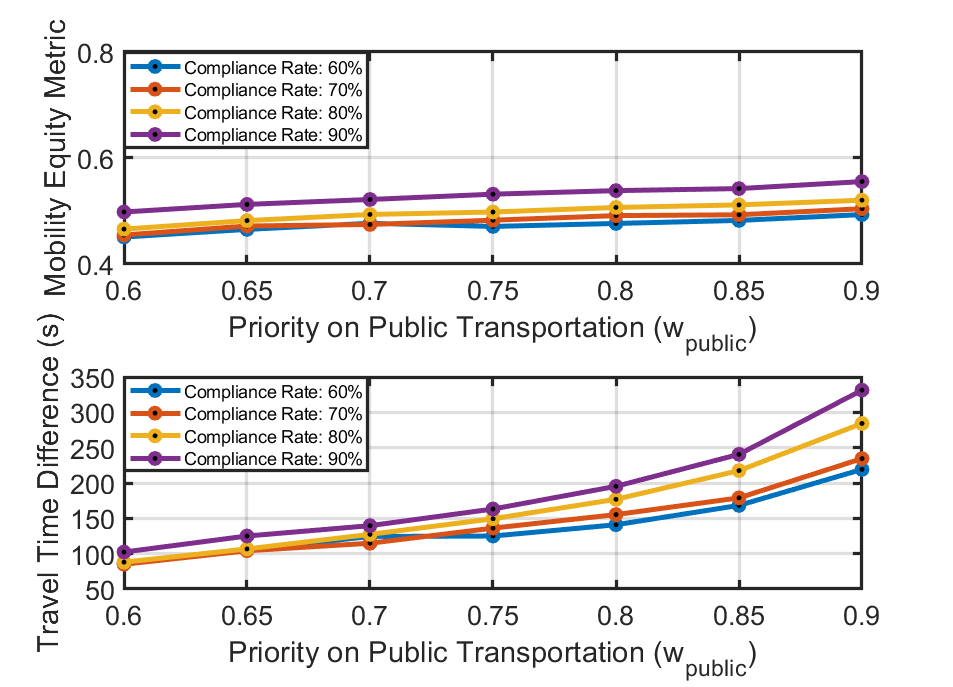}
             \caption{Percentage of public transportation = 30 $\%$}
             \label{fig:mem_30a}
    \end{subfigure}
    \hfill
    \begin{subfigure}[b]{0.49\textwidth}
        \includegraphics[width=\textwidth]{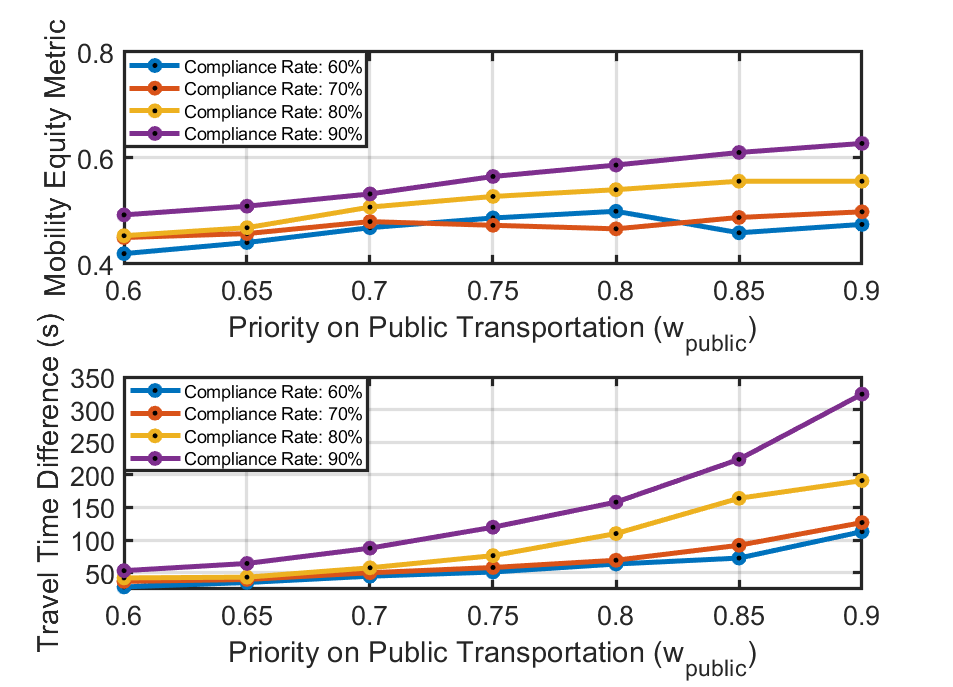}
             \caption{Percentage of public transportation = 50 $\%$}
             \label{fig:mem_30b}
    \end{subfigure}
        \caption{Evaluation of MEM and travel time difference with 3 hierarchy levels.}
    \label{fig:mem_3}
\end{figure*} 


\begin{figure*}[h!]
    \centering
    \begin{subfigure}[b]{0.49\textwidth}
        \includegraphics[width=\textwidth]{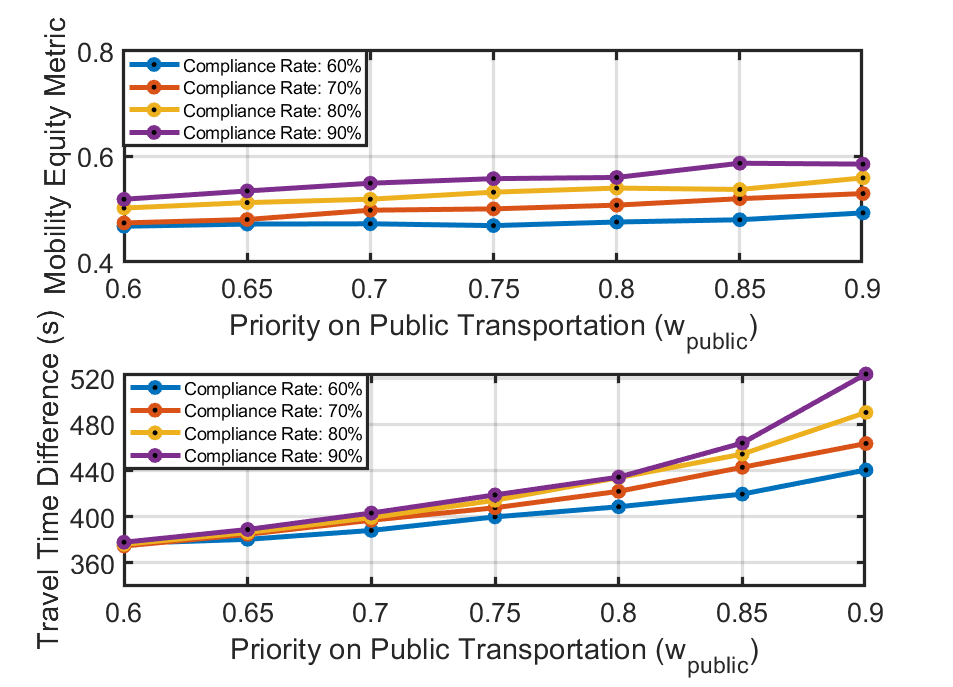}
             \caption{Percentage of public transportation = 30 $\%$}
             \label{fig:mem_wardrop_b}
    \end{subfigure}
    \hfill
    \begin{subfigure}[b]{0.49\textwidth}
        \includegraphics[width=\textwidth]{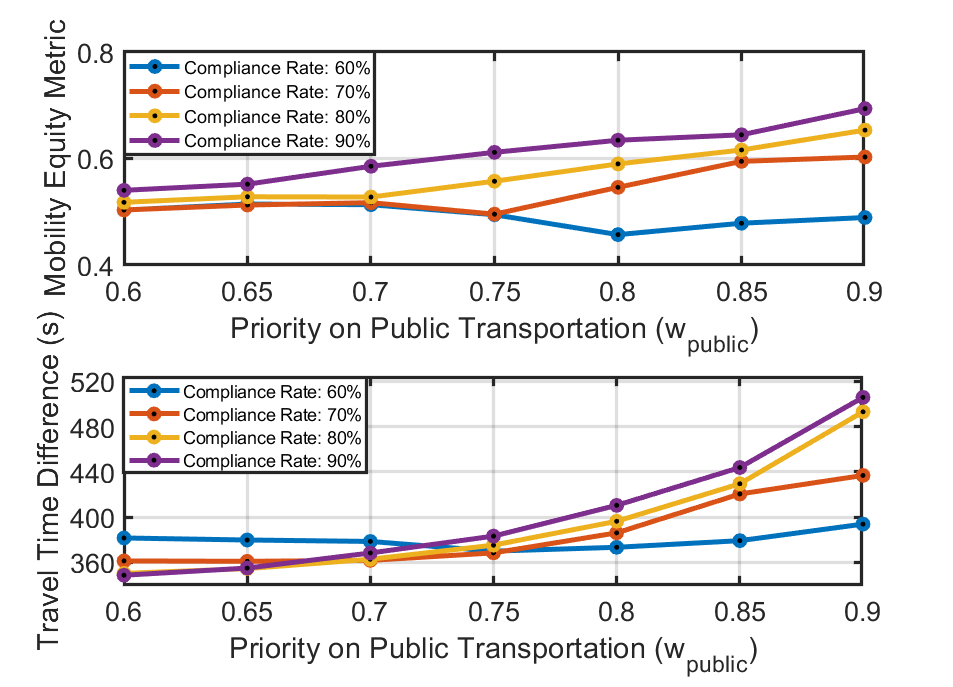}
             \caption{Percentage of public transportation = 50 $\%$}
             \label{fig:mem_wardrop_a}
    \end{subfigure}
        \caption{Evaluation of MEM and travel time difference using Wardrop Equilibrium for noncompliant vehicles.}
        \label{fig:mem_wardrop}
\end{figure*}

\begin{figure}
    \centering
    \includegraphics[width=1\linewidth]{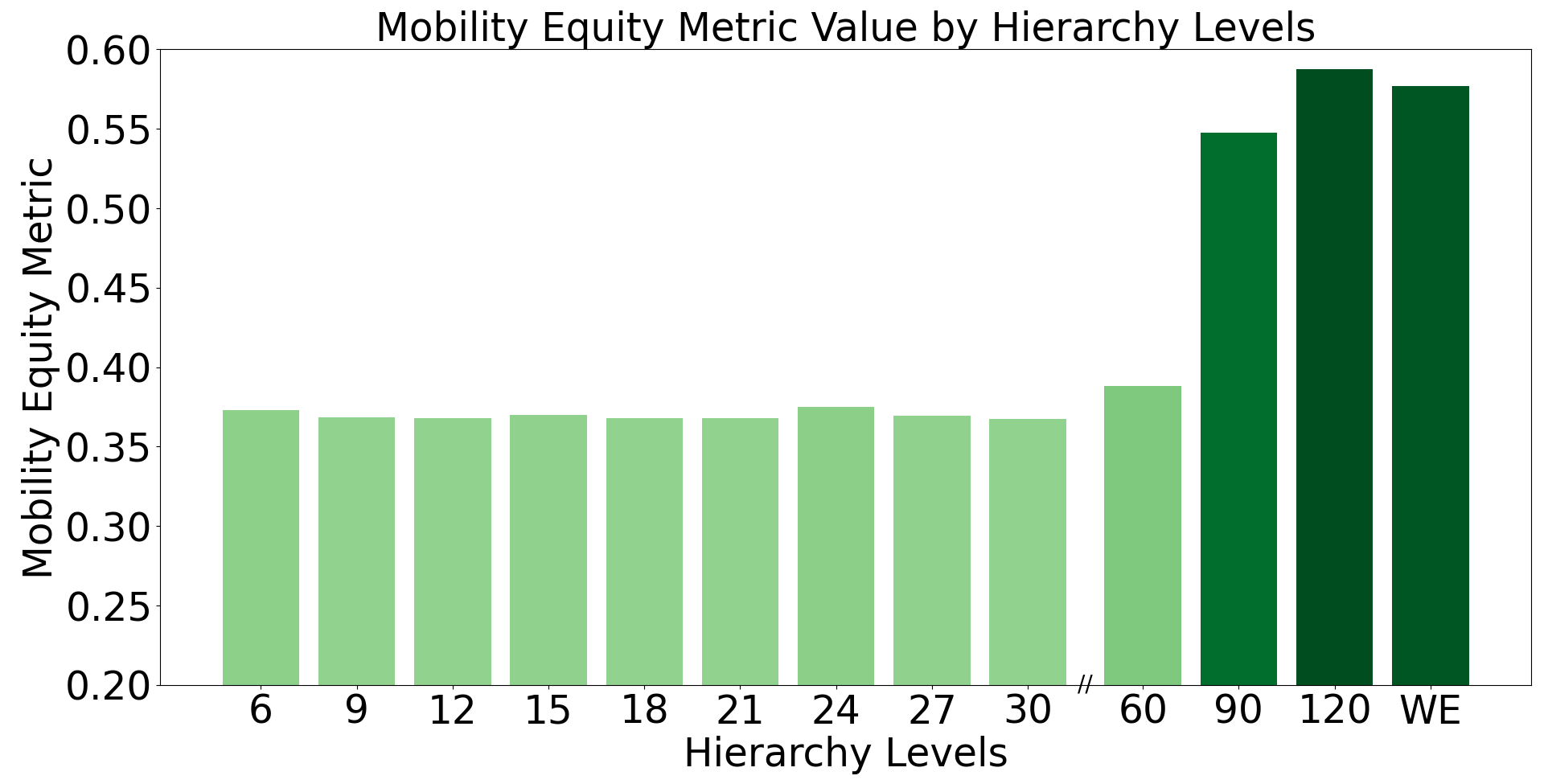}
    \caption{MEM values for different hierarchy levels and Wardrop Equilibrium.}
    \label{fig:small_hierarchy}
\end{figure}

\begin{figure} 
    \centering
    \includegraphics[width=1\linewidth]{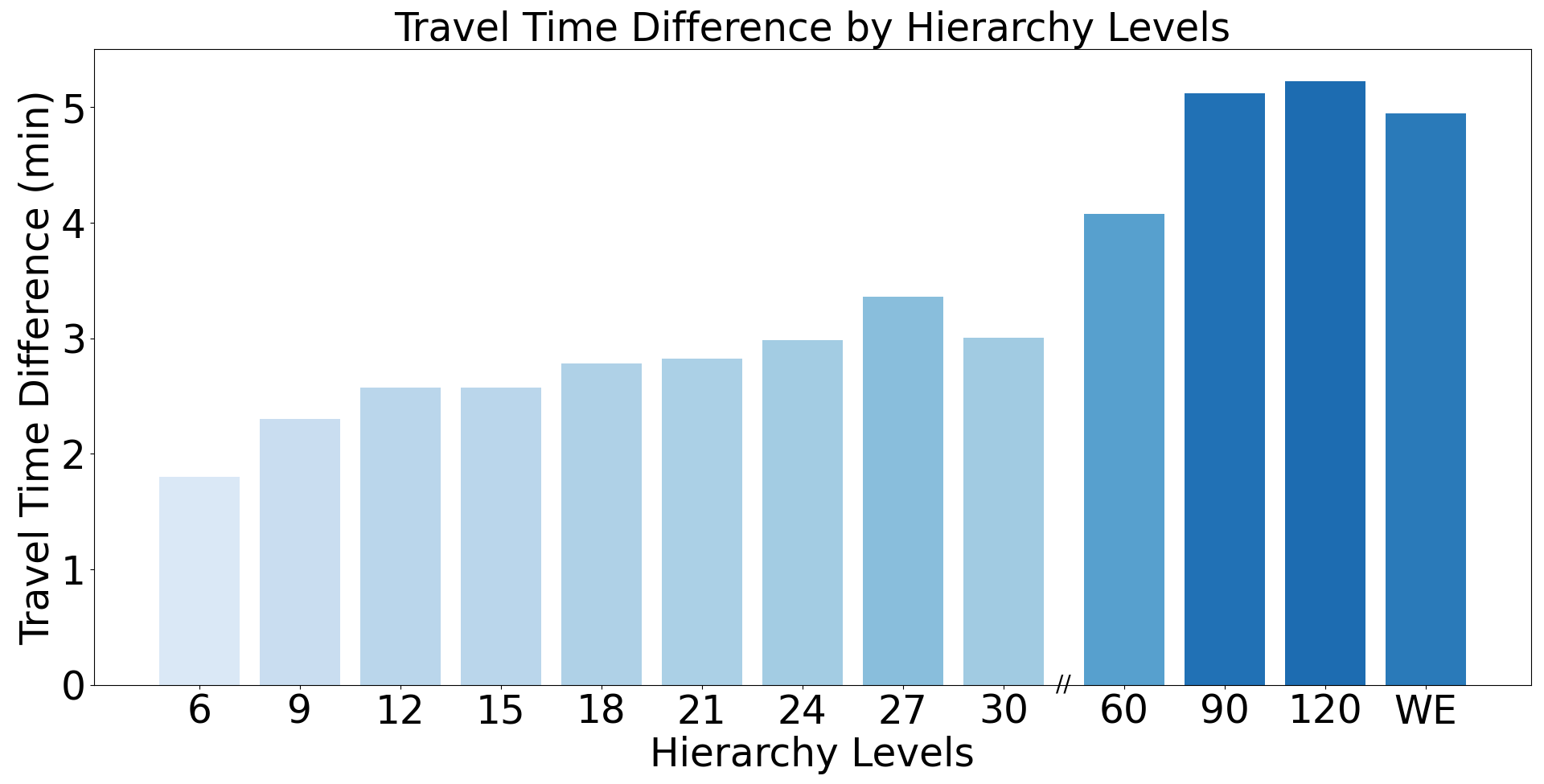}
    \caption{Travel time difference between compliant and noncompliant vehicles for different hierarchy levels and Wardrop equilibrium.}
    \label{fig:larger_hierarchy}
\end{figure}

\begin{figure}[h!]
    \centering
    \includegraphics[width=0.9\linewidth]{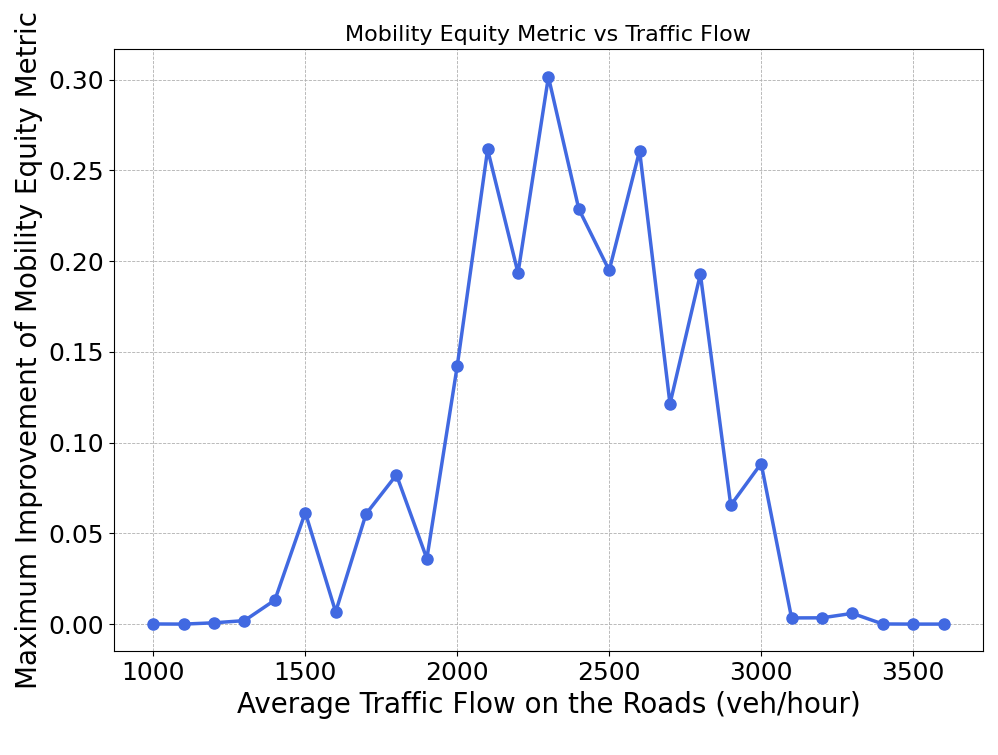}
    \caption{Maximum possible improvement of mobility equity metric among different compliance rates with respect to traffic flows in the network.}
    \label{fig:mem_improvement}
\end{figure}

\subsection{Simulation Results}
We initially restricted our model to level-2 decision-making for non-compliant vehicles, based on the experimental evidence \cite{costa2009comparing,costa2006cognition} suggesting that humans are rarely capable of reasoning beyond level-2 in cognitive hierarchy models.
Figure \ref{fig:mem_3} demonstrates the MEM values and the average travel time differences between compliant and non-compliant vehicles for different weights on public transportation.
We compared the results of our cognitive hierarchy model with simulations using the Wardrop equilibrium model for non-compliant vehicles (see Fig. \ref{fig:mem_wardrop}).
Wardrop equilibrium describes a state in traffic networks where all users have chosen routes that minimize their travel time, and no one can benefit by unilaterally switching routes. In our routing game, this implies that travel times on all paths are equal and optimal from the perspective of non-compliant drivers.

In both the cognitive hierarchy and Wardrop equilibrium scenarios, we observed increased MEM values as the weight on public transportation and compliance rates rose. This increase in MEM occurs because public transportation and non-compliant vehicles benefit from shorter travel times at the expense of compliant private vehicles' travel time. At the same time, the travel time difference between compliant and non-compliant vehicles widened since prioritizing public transportation resulted in longer travel times for compliant private vehicles. In other words, compliant private vehicles sacrifice travel time as the system prioritizes public transportation. 

Additionally, by comparing Figs. \ref{fig:mem_30a} and \ref{fig:mem_30b}, we observe that MEM increases more substantially when the proportion of public transportation is $50 \%$, as opposed to when the proportion is $30 \%$. This is reasonable, as prioritizing public transportation has a larger impact on the network when there are more public vehicles. The same trend can be seen in Figs. \ref{fig:mem_wardrop_a}
 and \ref{fig:mem_wardrop_b}.

\subsection{Impact of Hierarchy Levels}

We further explored the impact of different levels in the cognitive hierarchy model, as shown in Figs. \ref{fig:small_hierarchy} and \ref{fig:larger_hierarchy}. The results indicated that MEM increased as the hierarchy level increased. However, improvement only became noticeable beyond level $30$, beyond the realistic human decision-making capabilities.
The results also showed increased travel time differences between compliant and non-compliant vehicles as the hierarchy level increased. This suggests that non-compliant vehicles strategically chose less congested paths than compliant vehicles yielded to public transportation. The lack of ability to anticipate actual traffic conditions led to increased congestion and hindered the achievement of better mobility equity.

\subsection{Impact of Traffic Flows}
Lastly, we investigated potential MEM improvements with respect to different traffic flows in the network. Figure \ref{fig:mem_improvement} shows the maximum difference in MEM for compliance rates ranging from $60\%$ to $90\%$. The potential improvement in MEM increased as the traffic flow on the roads increased to $2300$ vehicles per hour. This is because higher traffic volumes provided more opportunities for the routing framework to optimize routes and improve mobility equity.
Meanwhile, for the flow beyond this point, MEM decreased due to traffic congestion, lowering the efficiency of the system planner's routes.

Overall, our simulations demonstrated the effectiveness of the proposed routing framework in enhancing mobility equity in the Boston metropolitan network. The results highlighted the importance of considering human drivers' cognitive limitations and the potential benefits of higher compliance rates and traffic volumes in achieving more equitable mobility outcomes.

\section{Concluding Remarks} \label{sec:conclusion}

We introduced MEM to evaluate fairness and accessibility in multi-modal intelligent transportation systems. MEM offers a comprehensive measure of mobility equity by incorporating travel time, user cost, price sensitivity, transportation modes, and service types. We provided a data-driven validation of the MEM across 12 major U.S. cities to demonstrate its effectiveness in capturing and comparing mobility equity across different urban environments. Subsequently, we developed a routing framework to optimize MEM within transportation networks containing emerging mobility systems and a social planner. In the routing framework, we accommodated both public transit and private vehicles as modes of transportation and allowed private vehicles to be either compliant or non-compliant with routing recommendations. By employing a cognitive hierarchy model, we captured the bounded rationality of human drivers more accurately than traditional Wardrop equilibrium approaches. We also provide a comparison with the Wardrop equilibrium-based model. Simulations based on the Boston Metropolitan area demonstrated the potential of emerging mobility systems in this framework to improve mobility equity through intelligent routing strategies.

The benefits of our work are multifaceted. First, MEM can be a valuable tool for urban planners and policymakers to quantify and compare mobility equity across different cities or neighborhoods, enabling data-driven decision-making in transportation planning. Second, our routing framework offers a practical approach to improving mobility equity in existing transportation networks without requiring significant infrastructure changes. This is particularly valuable given the constraints of urban space and the high costs associated with major infrastructure projects.

However, it is important to acknowledge the limitations of our approach, which opens up compelling possibilities for future work. Our current model does not allow for mode transfers within a single trip, thus failing to capture the complexity of multi-modal journeys in real-world scenarios. Incorporating this important consideration would make the routing problem computationally challenging and improve the applicability of the work.
Furthermore, neither the cognitive hierarchy framework nor the Wardrop equilibrium model may capture the complexities of human decision-making in routing. It is important to consider the integration of learned models for human behavior into the routing problem.
Finally, while our routing framework can improve mobility equity under a range of demands and network conditions, further significant improvements in equity often require interventions at the urban planning level. The strategic placement of essential services and the development of new transportation infrastructure. Further research is required to uncover the most impactful and cost-efficient approaches to improve mobility equity across different cities.

\section*{Acknowledgments}

The authors would like to acknowledge Songyang Ruan for helping with the data analysis in Subsection \ref{subsection:data analysis}, developing some figures, and developing the website for a live demonstration.


\bibliographystyle{IEEEtran}
\bibliography{Bang,IDS,MEM}

\end{document}